\documentclass[%
preprint, 
11pt, 
 amsmath,amssymb,
 aps, physrev,
]{revtex4-2}

\usepackage{graphicx}
\usepackage{dcolumn}
\usepackage{bm}

\usepackage{subcaption}
\usepackage{tikz,pgfplots,siunitx}
\usepackage{pgfplotstable}
\usepackage{amsmath}
\pgfplotsset{compat=1.18}

\def\tr[#1]{\textcolor{red}{#1}}
\def\tg[#1]{\textcolor{green}{#1}}
\def\tb[#1]{\textcolor{blue}{#1}}

\def\Rey{{\mathrm{Re}}}
\newcommand\Real{\mbox{Re}}          

\newcommand{\bff}{{\boldsymbol{f}}}

\newcommand{\bfn}{{\boldsymbol{n}}}

\newcommand{\bfq}{{\boldsymbol{q}}}

\newcommand{\bfR}{{\boldsymbol{R}}}

\newcommand{\bfu}{{\boldsymbol{u}}}

\newcommand{\babfu}{\overline{\bfu}}
\newcommand{\babfq}{\overline{\bfq}}

\newcommand{\hbfq}{\boldsymbol{\hat{q}}}

\newcommand{\hbff}{\boldsymbol{\hat{f}}}
\newcommand{\hbfu}{\boldsymbol{\hat{u}}}
\newcommand{\hbfv}{\boldsymbol{\hat{v}}}

\newcommand{\hbfPsi}{\boldsymbol{\hat{\Psi}}}

\renewcommand{\leq}{\leqslant}
\renewcommand{\geq}{\geqslant}

\makeatletter
\newcommand{\eqautoref}[1]{%
  \begingroup
  \def\equationautorefname~##1\null{%
    \ifnum\count@=1
      Eq.~(##1)%
    \else
      Eqs.~(##1)%
    \fi
    \null
  }%
  \eqautoref@parse#1,\relax
  \endgroup
}
\def\eqautoref@parse#1,#2\relax{%
  \count@=0
  \eqautoref@item{#1}%
  \ifx\\#2\\%
  \else
    ,\eqautoref@parse#2\relax
  \fi
}
\def\eqautoref@item#1{%
  \advance\count@ by 1
  \autoref{#1}%
}
\makeatother

\begin{document}

\preprint{APS/123-QED}

\title{\textbf{Coherent structures modeling in stenotic transitional flow via resolvent analysis} 
}
 
\author{A. Villié}
\email{Contact author: a.villie@tu-berlin.de}
\author{S. Demange}%
\author{K. Oberleithner}
\affiliation{%
 Laboratory for Flow Instability and Dynamics, Technische Universität Berlin, 10623 Berlin, Germany.}%

\date{\today}

\begin{abstract}
This study investigates the capability of linear modeling to characterize the transitional dynamics in an axisymmetric stenosis and attempts a low-order representation of the turbulent stresses. 
The transition to turbulence in stenotic flows generates wall shear stress fluctuations that strongly influence the progression of cardiovascular diseases and the risk of plaque rupture. A description of the linear mechanisms driving the forced dynamics at Reynolds number beyond transition is currently missing. 
Linear modeling of coherent structures is leveraged to identify the flow amplification mechanisms using the mean field from a large-eddy simulation at a Reynolds number of $4000$.
Global linear stability analysis reveals an unstable and sinuous stationary eigenmode that is known to destabilize the flow at lower Reynolds numbers through a weak Coanda-type wall attachment. 
Resolvent analysis confirms the dominance of this stationary mode at low frequencies. 
At intermediate frequencies, it identifies a second amplification region within the shear-layer where the most amplified fluctuations are axisymmetric, in contrast to findings from studies at lower Reynolds numbers.
The linear model is validated against spectral proper orthogonal decomposition (SPOD). At intermediate frequencies, the optimal resolvent response mode demonstrates both high gain separation and strong alignment with the leading SPOD mode. 
The low-rank nature of the resolvent operator is leveraged to reconstruct the turbulent kinetic energy (TKE) and turbulent wall shear stress (tWSS) from the optimal response mode. 
In the immediate post-stenotic zone, axisymmetric fluctuations dominate the tWSS and exhibit low-rank dynamics, with the leading SPOD mode capturing more than half of the tWSS.
Our findings highlight that linear mechanisms effectively capture the complex post-stenotic dynamics, providing physically interpretable insights into the flow physics.
The successful reconstruction of turbulent quantities from mean flow data alone opens new predictive possibilities of key turbulent quantities. 
\end{abstract}
\maketitle

\section{Introduction}
The dynamics of blood flow past arterial stenoses play a central role in the development and progression of cardiovascular diseases. 
Atherosclerosis, considered the most fatal of these pathologies, is characterized by the accumulation of plaque in the inner layer of arteries. This induces a localized narrowing, called stenosis, which can lead to non-recoverable pressure loss and flow choking, significantly increasing the risk of heart attack or stroke.
As the flow accelerates through the constriction, it induces high wall shear stress (WSS) at the stenosis throat. These elevated stresses promote platelet activation and thrombosis, potentially leading to plaque rupture and vessel occlusion. Conversely, the downstream disturbed flow regions exhibit low-magnitude and oscillatory shear stresses, which have been associated with arterial wall thickening and the progression of atherosclerotic disease \cite{ku1997blood, peiffer2013does}.

Hemodynamics also impact the progression of cardiovascular diseases through flow-sensitive endothelial cells.
Under physiological laminar flow conditions, these cells align along the direction of flow and maintain vascular integrity \cite{davies2009hemodynamic}. When transition to turbulence occurs, the chaotic WSS disturbs this alignment, increasing susceptibility to vascular diseases \cite{ziegler2017assessment}. More specifically, low-amplitude, oscillatory, and multidirectional WSS conditions have been directly associated with endothelial dysfunction \cite{peiffer2013does, andersson2017multidirectional}.
Additionally, a number of studies propose that hemodynamics influence diseases indirectly by altering the transport of solutes—such as oxygen or low-density lipoproteins—in the near-wall fluid, rather than through direct mechanical effects on the wall itself \cite{ma1997convective}. 
Beyond endothelial dysfunction, transitional flow may also induce mechanical fatigue in the stenotic plaque \cite{ojha1994wall}. The turbulent stresses acting on an atherosclerotic plaque may promote high shear stresses, pressure fluctuations and erosion, which may cause plaque rupture \cite{stroud2002numerical, slager2005role} and lead to distal embolisms. 
Therefore, detailed characterization of the WSS dynamics is essential to improve diagnostic accuracy and inform more effective treatment strategies.

Experimental studies at physiologically relevant Reynolds numbers demonstrate that the onset of unsteadiness and transition to turbulence in post-stenotic flow is driven by vortex shedding from the constriction throat \cite{ahmed1983flow}. 
To quantify the WSS fluctuations produced by these large vortices, \citet{lantz2012quantifying} introduced the turbulent wall shear stress (tWSS) as a biomarker to characterize the WSS unsteadiness.
The tWSS directly correlates with near-wall turbulent kinetic energy (TKE) \cite{ziegler2017assessment, andersson2017multidirectional}, and can reach up to 40\% of the total WSS in aortic valve stenosis, highlighting the importance of WSS fluctuations \cite{manchester_analysis_2021}.

Over the last two decades, computational fluid dynamics (CFD) has become a primary tool for computing time-resolved flow data. Although CFD enables high-resolution simulations in complex geometries, it typically represents the flow as a series of temporal snapshots, which does not directly reveal the dominant coherent structures.
Developments in computational and experimental fluid mechanics have led to the generation of increasingly large volumes of flow data. However, for complex configurations, such as cardiovascular flows, extracting physical insight from the highly resolved spatio-temporal datasets remains challenging.
To facilitate the understanding and modeling of turbulent flows, one can seek a basis to express the dynamics as a collection of modes ranked by energy content or frequency.
Recent studies have increasingly turned to data-driven reduced-order modeling approaches to characterize physiological flows \cite{csala2025exploiting}. For instance, Proper Orthogonal Decomposition (POD) has been used to visualize transitional flow patterns in stenosed carotid arteries \cite{kefayati2013transitional}, to reconstruct WSS in dysfunctional mechanical heart valves \cite{darwish2021proper},
and to extract energetic modes in severely stenosed arteries \cite{kazemi2022reduced}.
Dynamic Mode Decomposition (DMD) has been applied to brain aneurysm flows to analyze instabilities \cite{le_dynamic_2021, nguyen2025modal} and patient-specific cardiovascular flow data to extract coherent WSS structures \cite{habibi2020data}.
Finally, Spectral Proper Orthogonal Decomposition (SPOD) combines the energy-ranking of POD and the frequency correlation of DMD, providing a powerful spectral framework to identify the most energetic spatiotemporal coherent patterns \cite{towne_spectral_2018}.
It has recently been used in cerebral aneurysms \cite{luciano2025decomposition} and abdominal aortic aneurysms \cite{norouzi2021flow} to identify the coherent structures. 

While data-driven methods reveal flow organization, they often lack physical interpretability and are unable to predict unresolved features.
In addition, they rely on large, highly-resolved datasets since the spectral resolution is constrained by the total signal duration, leading to substantial computational costs. This dependency on well-resolved datasets limits their applicability to clinical measurement techniques, such as 4D-flow magnetic resonance imaging (4D-flow MRI), where temporal and spatial resolution are constrained \cite{markl20124d}.

An alternative to data-driven approaches is to rely on physics-based models.
By linearizing the Navier-Stokes equations about a base flow, one can investigate the linear mechanisms driving the growth/decay of coherent structures, eliminating the need for time-resolved data.
Such linear stability analysis (LSA) reveals the linear modal mechanisms of the unforced system \cite{schmid2001stability}. 
LSA of the steady base flow through an axisymmetric arterial stenosis has first served as an idealized physics-based model for understanding the laminar-turbulent transition mechanisms. \citet{sherwin2005three} showed that the 75\% area-reduction stenosis with steady inflow becomes unstable above a critical Reynolds number of $\Rey_c = 722$. 
The leading eigenmode is stationary and 
corresponds to the azimuthal wavenumber $m=1$. This mode causes jet deflection and asymmetric attachment, a phenomenon described as 'mild Coanda-type attachment'.
However, experimental studies have observed flow unsteadiness significantly below the global instability threshold at $\Rey_c$ \cite{khalifa1981characterization, ahmed1983velocity, vetel2008asymmetry}. 
Direct numerical simulation (DNS) successfully reproduced transition consistent with experimental observations when a small geometric asymmetry or inflow swirl is introduced \cite{varghese2007direct}. This comparison strongly suggests that transition in experiments is mainly noise-amplified rather than governed by a modal global instability.

The early transition observed in experiments was later explained by \citet{blackburn2008convective} and \citet{griffith2010convective} using optimal transient growth analysis. They showed that small perturbations in the post-stenotic shear-layer can amplify by several orders of magnitude through convective instability. For severe stenosis cases, this process manifests as shear-layer roll-up into coherent axisymmetric vortices, which subsequently lose symmetry and break down into turbulence \cite{ojha1994wall, varghese2007direct, han2025lagrangian}. This convective instability has since been established as the principal route to transition in realistic stenotic flows \cite{griffith2008steady, griffith2010convective}. 
However, while transient growth analysis focuses on initial-value problems, it does not characterize persistent flow structures arising from continuous nonlinear forcing, observed beyond the critical Reynolds number in turbulent conditions \cite{vetel2008asymmetry}. 
A comprehensive description of the energetic distribution of these coherent structures is currently lacking.

In recent years, resolvent analysis (RA) has become widely used to model coherent structures in amplifier-type flows.
By retaining the nonlinear terms as a source of external forcing, it enables an input-output description of the transient (non-modal) amplification mechanisms \cite{mckeon2010critical}. 
When the mean turbulent solution is used as the base flow, it reveals dominant linear mechanisms, even in the absence of any amplified global mode. In recent years, RA has been employed to identify the physical processes that form large-scale coherent structures in a variety of turbulent flows, including turbulent jets~\cite{kuhn2021linear, pickering_optimal_2021}, flows around airfoils~\cite{symon_tale_2019, demange2024resolvent}, pipe flows~\cite{abreu2020spectral, Müller_2024}, and turbulent separation bubbles~\cite{cura2025linear, fuchs2025standing}. 
\citet{towne_spectral_2018} established a direct correspondence between the optimal RA response modes and the SPOD modes when the true nonlinear forcing is white noise, allowing validation of the resolvent model using empirical data-driven methods.

When properly calibrated, resolvent modes can successfully reconstruct transitional flows \cite{symon2020mean} and provide low-order representations of the linear amplification mechanisms.
This capability to reconstruct dynamics solely from the mean state makes the RA particularly attractive for clinical measurement techniques with limited temporal resolution, like 4D-flow Magnetic Resonance Imaging (4D-flow MRI \cite{markl20124d}).
Such RA-based models have been successfully applied to a wide range of configurations, including turbulent channel flows \cite{symon2018non}, jet flows \cite{oberleithner_three-dimensional_2011, kuhn2021linear, towne_spectral_2018}, flow around an airfoil \cite{thomareis2018resolvent, yeh2019resolvent, demange2024resolvent}, and separation bubbles \cite{yeh2020resolvent, fuchs2025standing, cura2025linear}. 
However, to the authors' knowledge, it has yet to be applied to cardiovascular flows.

This work focuses on a stenotic flow at Reynolds number $4000$ with a $75\%$ area reduction. The objectives are twofold. 
First, we want to characterize the physical mechanisms affecting the flow dynamics.
To this end, LES is used to generate high-fidelity flow data from which SPOD modes are obtained, while a physics-based model of the coherent structures, constructed from the mean flow, is employed to identify the dominant flow dynamics.
Second, we assess the ability of a low-order model to reproduce the TKE and tWSS. Specifically, we examine (i) the contribution of dominant coherent structures, across different azimuthal wavenumbers and frequencies, to the TKE and tWSS and (ii) the ability of RA to reconstruct the turbulent stresses using only the time-averaged mean flow.
The paper is organized as follows.
Section~\ref{sec:numerical_simulation} introduces the numerical setup and the main flow characteristics. 
Section~\ref{sec:linModel} presents the linear stability framework. To fully characterize the amplification mechanisms, we perform LSA and RA, and analyse the pseudospectrum. 
Section~\ref{sec:data_driven} validates the linear model with SPOD and discusses the white noise forcing assumption.
Section~\ref{sec:ROM} builds a low-order model of the RST and tWSS from the leading resolvent and SPOD modes. 
Finally, conclusions are drawn in section~\ref{sec:Conclusion}.

\section{Numerical simulation}\label{sec:numerical_simulation}
\begin{figure}%
    \centering
    \def\svgwidth{0.8\textwidth}
    \fontsize{12}{1}\selectfont
\begingroup%
  \makeatletter%
  \providecommand\color[2][]{%
    \errmessage{(Inkscape) Color is used for the text in Inkscape, but the package 'color.sty' is not loaded}%
    \renewcommand\color[2][]{}%
  }%
  \providecommand\transparent[1]{%
    \errmessage{(Inkscape) Transparency is used (non-zero) for the text in Inkscape, but the package 'transparent.sty' is not loaded}%
    \renewcommand\transparent[1]{}%
  }%
  \providecommand\rotatebox[2]{#2}%
  \newcommand*\fsize{\dimexpr\f@size pt\relax}%
  \newcommand*\lineheight[1]{\fontsize{\fsize}{#1\fsize}\selectfont}%
  \ifx\svgwidth\undefined%
    \setlength{\unitlength}{219.81998954bp}%
    \ifx\svgscale\undefined%
      \relax%
    \else%
      \setlength{\unitlength}{\unitlength * \real{\svgscale}}%
    \fi%
  \else%
    \setlength{\unitlength}{\svgwidth}%
  \fi%
  \global\let\svgwidth\undefined%
  \global\let\svgscale\undefined%
  \makeatother%
  \begin{picture}(1,0.31559913)%
    \lineheight{1}%
    \setlength\tabcolsep{0pt}%
    \put(0,0){\includegraphics[width=\unitlength,page=1]{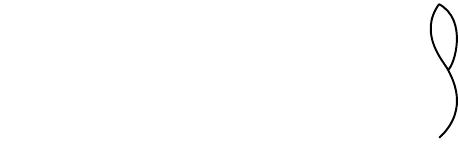}}%
    \put(0.62386474,0.15147037){\color[rgb]{0,0,0}\makebox(0,0)[lt]{\lineheight{0}\smash{\begin{tabular}[t]{l}$x$\end{tabular}}}}%
    \put(0.47371311,0.19453995){\color[rgb]{0,0,0}\makebox(0,0)[lt]{\lineheight{0}\smash{\begin{tabular}[t]{l}$r$\end{tabular}}}}%
    \put(0.52504198,0.20537249){\color[rgb]{0,0,0}\makebox(0,0)[lt]{\lineheight{0}\smash{\begin{tabular}[t]{l}$\theta$\end{tabular}}}}%
    \put(0,0){\includegraphics[width=\unitlength,page=2]{Expsetup.pdf}}%
    \put(0.30846708,0.18587927){\color[rgb]{0,0,0}\makebox(0,0)[lt]{\lineheight{0}\smash{\begin{tabular}[t]{l}$d_t$\end{tabular}}}}%
    \put(0,0){\includegraphics[width=\unitlength,page=3]{Expsetup.pdf}}%
    \put(0.00000118,0.15538353){\color[rgb]{0,0,0}\makebox(0,0)[lt]{\lineheight{0}\smash{\begin{tabular}[t]{l}$D$\end{tabular}}}}%
    \put(0,0){\includegraphics[width=\unitlength,page=4]{Expsetup.pdf}}%
  \end{picture}%
\endgroup%

    \caption{Sketch of the geometry.}
    \label{fig:exp_setup}
\end{figure}

We investigate the dynamics of an axisymmetric stenosis using LES. The flow configuration, illustrated in figure~\ref{fig:exp_setup}, consists of a cosine-shaped stenosis where the unobstructed diameter $D = 0.019$ m reduces to $d_t = D/2$, corresponding to a $75\%$ area reduction (severe stenosis).
A cylindrical coordinate system $(x, r, \theta)$ aligned with the stenosis axis is employed to describe the flow.

The fluid is modeled as incompressible and Newtonian with kinematic viscosity $\nu = 2.7 \times 10^{-6}$ m$^2$/s . This Newtonian assumption is appropriate for large-diameter vessels where shear rates are large \cite{pedley1980fluid, jin2003effects}. Similarly, we assume rigid walls, as compliance effects are generally negligible in larger arteries \cite{ku1997blood}.
At the inlet, we prescribe a steady laminar parabolic velocity profile with bulk velocity $U=0.57$ m/s, corresponding to an inlet Reynolds number of $\Rey=4000$. This is representative of peak systolic conditions in the aorta \cite{ku1997blood}. 
A constant pressure is set at the outlet, and no-slip conditions are applied at the walls. 

Experimentally observed transition to turbulence in the post-stenotic region is dominated by the characteristics of the constriction rather than the main vessel \cite{cassanova1978disorder, deshpande1980turbulence, ahmed1983flow}.
Consequently, we characterize the dimensionless frequencies using a Strouhal number based on the throat parameters:
\begin{equation}
    \mathrm{St} = \dfrac{f d_t}{U_t},
\end{equation}
where $f$ is the dimensional frequency and $U_t = 4U$ is the bulk velocity at the stenosis throat.
All subsequent results and variables are normalized using the obstructed diameter $d_t$ and throat velocity $U_t$.

\begin{figure}%
    \centering
    \def\svgwidth{0.8\textwidth}
\begingroup%
  \makeatletter%
  \providecommand\color[2][]{%
    \errmessage{(Inkscape) Color is used for the text in Inkscape, but the package 'color.sty' is not loaded}%
    \renewcommand\color[2][]{}%
  }%
  \providecommand\transparent[1]{%
    \errmessage{(Inkscape) Transparency is used (non-zero) for the text in Inkscape, but the package 'transparent.sty' is not loaded}%
    \renewcommand\transparent[1]{}%
  }%
  \providecommand\rotatebox[2]{#2}%
  \newcommand*\fsize{\dimexpr\f@size pt\relax}%
  \newcommand*\lineheight[1]{\fontsize{\fsize}{#1\fsize}\selectfont}%
  \ifx\svgwidth\undefined%
    \setlength{\unitlength}{107.58301262bp}%
    \ifx\svgscale\undefined%
      \relax%
    \else%
      \setlength{\unitlength}{\unitlength * \real{\svgscale}}%
    \fi%
  \else%
    \setlength{\unitlength}{\svgwidth}%
  \fi%
  \global\let\svgwidth\undefined%
  \global\let\svgscale\undefined%
  \makeatother%
  \begin{picture}(1,0.24188454)%
    \lineheight{1}%
    \setlength\tabcolsep{0pt}%
    \put(-0.0003024,0.2170005){\color[rgb]{0,0,0}\makebox(0,0)[lt]{\lineheight{0}\smash{\begin{tabular}[t]{l}a)\end{tabular}}}}%
    \put(-0.00061921,0.09972464){\color[rgb]{0,0,0}\makebox(0,0)[lt]{\lineheight{0}\smash{\begin{tabular}[t]{l}b)\end{tabular}}}}%
    \put(0.76851886,0.09972464){\color[rgb]{0,0,0}\makebox(0,0)[lt]{\lineheight{0}\smash{\begin{tabular}[t]{l}c)\end{tabular}}}}%
    \put(0,0){\includegraphics[width=\unitlength,page=1]{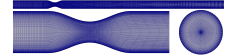}}%
  \end{picture}%
\endgroup%

    \caption{Axisymmetric structured mesh. A 2D slice of full domain (a) is displayed with a zoom on the stenosis (b) and a circular cross-section (c).}
    \label{fig:mesh}
\end{figure}
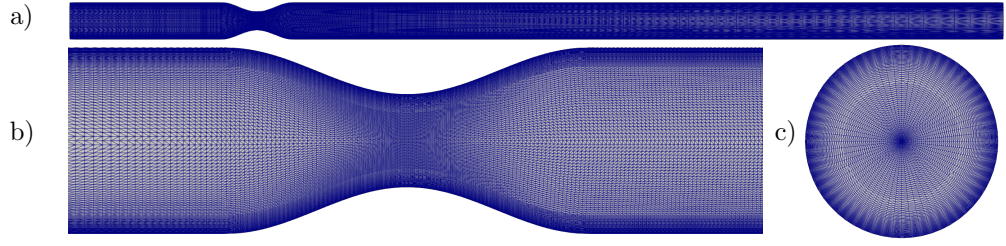

The incompressible Navier-Stokes equations were solved using the \texttt{pimpleFoam} solver in OpenFOAM v2212. For turbulence modeling, the model $k-\omega$ shear-stress transport detached-eddy simulation ($k-\omega$ SST DES) was used to model dissipation from sub-grid scale structures \cite{strelets2001detached}. A similar numerical setup has been used by \citet{dillinger2022fundamentals} to study the turbulent spectrum. 
Spatial discretization is achieved with an axisymmetric structured mesh, as displayed in figure~\ref{fig:mesh}.
A mesh convergence study was performed to ensure grid-independent results, and a mesh with $1.5 \times 10^6$ cells was retained. The boundary layer is carefully resolved with $y^+<1$ everywhere, and the time step was set to $1.2 \times 10^{-2} \, d_t/U_t$ to maintain a Courant number below $1$. 

\begin{figure*}
    \centering
    \def\svgwidth{0.95\textwidth}
\begingroup%
  \makeatletter%
  \providecommand\color[2][]{%
    \errmessage{(Inkscape) Color is used for the text in Inkscape, but the package 'color.sty' is not loaded}%
    \renewcommand\color[2][]{}%
  }%
  \providecommand\transparent[1]{%
    \errmessage{(Inkscape) Transparency is used (non-zero) for the text in Inkscape, but the package 'transparent.sty' is not loaded}%
    \renewcommand\transparent[1]{}%
  }%
  \providecommand\rotatebox[2]{#2}%
  \newcommand*\fsize{\dimexpr\f@size pt\relax}%
  \newcommand*\lineheight[1]{\fontsize{\fsize}{#1\fsize}\selectfont}%
  \ifx\svgwidth\undefined%
    \setlength{\unitlength}{1097.61021892bp}%
    \ifx\svgscale\undefined%
      \relax%
    \else%
      \setlength{\unitlength}{\unitlength * \real{\svgscale}}%
    \fi%
  \else%
    \setlength{\unitlength}{\svgwidth}%
  \fi%
  \global\let\svgwidth\undefined%
  \global\let\svgscale\undefined%
  \makeatother%
  \begin{picture}(1,0.3788537)%
    \lineheight{1}%
    \setlength\tabcolsep{0pt}%
    \put(0,0){\includegraphics[width=\unitlength,page=1]{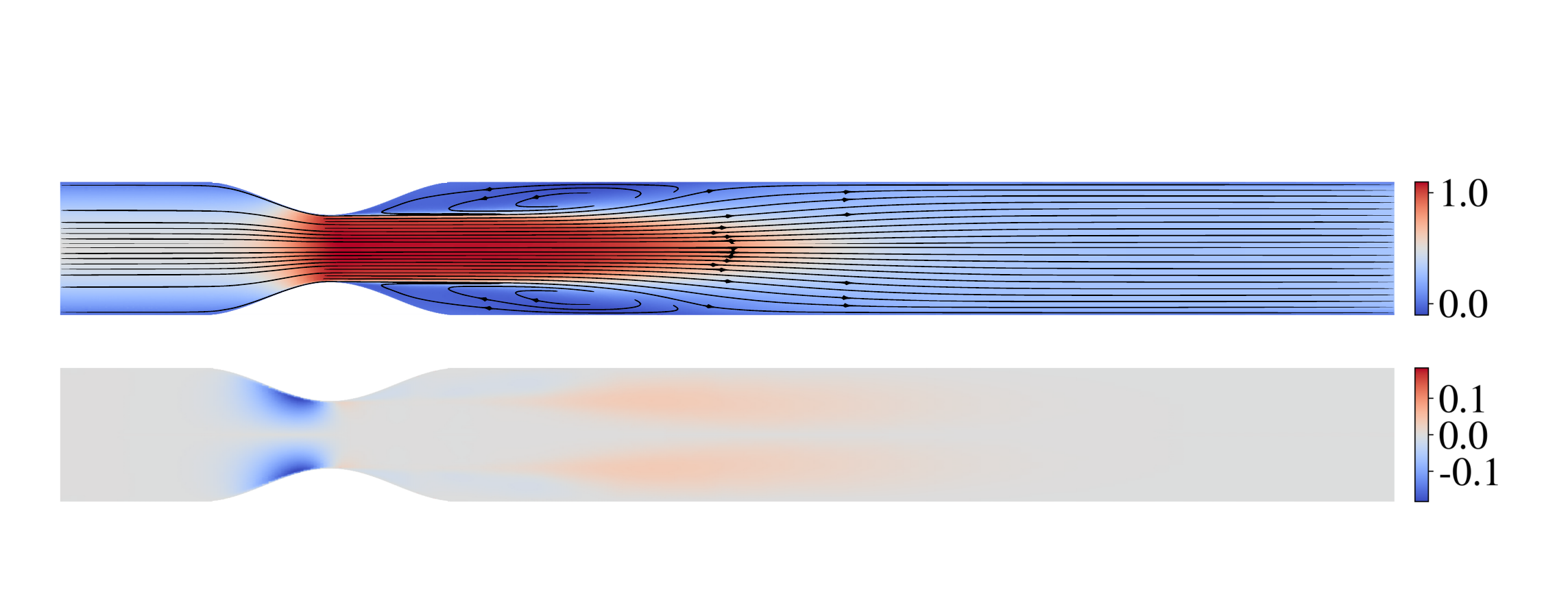}}%
    \put(0.96242153,0.21971828){\color[rgb]{0,0,0}\makebox(0,0)[lt]{\lineheight{0}\smash{\begin{tabular}[t]{l}$\overline{u}_x/U_t$\end{tabular}}}}%
    \put(0.96242153,0.09473064){\color[rgb]{0,0,0}\makebox(0,0)[lt]{\lineheight{0}\smash{\begin{tabular}[t]{l}$\overline{u}_r/U_t$\end{tabular}}}}%
    \put(0.47720957,0.00362163){\color[rgb]{0,0,0}\makebox(0,0)[lt]{\lineheight{0}\smash{\begin{tabular}[t]{l}$x/d_t$\end{tabular}}}}%
    \put(0,0){\includegraphics[width=\unitlength,page=2]{Q_criterion_mean_flow_streamlines.pdf}}%
    \put(-0.00043549,0.36467111){\color[rgb]{0,0,0}\makebox(0,0)[lt]{\lineheight{0}\smash{\begin{tabular}[t]{l}a)\end{tabular}}}}%
    \put(-0.00009742,0.1210276){\color[rgb]{0,0,0}\makebox(0,0)[lt]{\lineheight{0}\smash{\begin{tabular}[t]{l}c)\end{tabular}}}}%
    \put(-0.00068462,0.24558263){\color[rgb]{0,0,0}\makebox(0,0)[lt]{\lineheight{0}\smash{\begin{tabular}[t]{l}b)\end{tabular}}}}%
    \put(0,0){\includegraphics[width=\unitlength,page=3]{Q_criterion_mean_flow_streamlines.pdf}}%
  \end{picture}%
\endgroup%

    \caption{3D instantaneous contours of the Q-criterion clipped on a half plane (a); time- and azimuthal-average of the axial (b) and radial (c) velocity component, with superimposed streamlines.}
    \label{fig:Q_c_mean_flow}
\end{figure*}

Figure~\ref{fig:Q_c_mean_flow} shows the instantaneous contours of
the Q-criterion along with the temporal and azimuthal averaged velocity field from LES. 
Figure~\ref{fig:Q_c_mean_flow}(b) depicts a flow separation starting at the stenosis throat, forming a confined jet and an axisymmetric recirculation bubble that extends up to $x/d_t= 5.8$, where the jet reattaches.
As depicted by the Q-criterion iso-contours in figure~\ref{fig:Q_c_mean_flow}(a), axisymmetric rollers grow in the shear-layer ($x/d_t \in [0, 2.5]$).
These axisymmetric vortices then lose symmetry from $x/d_t=3$ and break down into 3D, chaotic turbulence. The development of these coherent structures and their breakdown to turbulence is typical in steady stenotic flows for $Re > 375$ \cite{bluestein1999vortex}. The immediate post-stenotic zone is a critical region to capture the stresses, since the recirculation bubble produces low and oscillatory WSS \cite{varghese2007direct}, potential driver of atherosclerotic disease progression \cite{ku1997blood, peiffer2013does} and plaque rupture \cite{ojha1994wall, stroud2002numerical}.

\section{Linear modeling} \label{sec:linModel}%
This section presents the linear analysis of the linear operator constructed about the mean flow, used as a physics-based model of the coherent fluctuations.
The 2D state vector $\bfq = [\bfu, p]^T$ contains the velocity field $\bfu$ in cylindrical coordinates and the pressure $p$. 
The Reynolds decomposition separates $\bfq$ into mean and fluctuating components, $\bfq = \babfq + \bfq'$.
The mean flow $\babfq$ is computed by time- and azimuthal-average of the LES snapshots [see figure~\ref{fig:Q_c_mean_flow}(b,c)].
Substituting this decomposition into the incompressible Navier-Stokes equations and subtracting the temporal mean yields the perturbation equations
\begin{eqnarray}
\label{eq:NS_perturbation1}
& \frac{\partial \bfu'}{\partial t}+(\babfu \cdot \boldsymbol{\nabla}) \bfu' + (\bfu' \cdot \boldsymbol{\nabla}) \babfu + \frac{1}{\rho}\boldsymbol{\nabla} p'- \boldsymbol{\nabla} \cdot \left[ \nu\,[\boldsymbol{\nabla}  + \boldsymbol{\nabla} ^T]\bfu' \right]= \bff_{nl}',\\
\label{eq:NS_perturbation2}
& \boldsymbol{\nabla} \cdot \bfu' = 0,
\end{eqnarray}
where $\rho$ is the mass density, $\nu$ is the kinematic viscosity, and $\bff_{nl}'$ contains the exact nonlinear terms, acting as a turbulent forcing. 

Rather than treating the nonlinear fluctuations purely as an external forcing, we apply a Boussinesq approximation using a RANS-like eddy viscosity, $\nu_t$, to relate the turbulent forces, $\boldsymbol{\nabla} \boldsymbol{\cdot} (\boldsymbol{u}^{\prime} \boldsymbol{u}^{\prime})$, to the fluctuating strain rate tensor. 
This Boussinesq modeling incorporates turbulent dissipation and is known to improve the linear analysis \cite{symon_use_2023, vonsaldernRoleEddyViscosity2024} and substantially increases the agreement between resolvent analysis and SPOD~\cite{pickering_optimal_2021, kuhn2022influence}. 
Introducing a frozen eddy viscosity (i.e. constant in time) modifies equation~\ref{eq:NS_perturbation1}:
\begin{eqnarray}
\label{eq:NS_perturbation1_mod}
& \frac{\partial \bfu'}{\partial t}+(\babfu \cdot \boldsymbol{\nabla}) \bfu' + (\bfu' \cdot \boldsymbol{\nabla}) \babfu + \frac{1}{\rho} \boldsymbol{\nabla} p'-\boldsymbol{\nabla} \cdot \left[ \left(\nu+\nu_t\right) \,[\boldsymbol{\nabla}  + \boldsymbol{\nabla} ^T]\bfu' \right] = \bff',
\end{eqnarray}
where $\bff'$ represents the remaining unmodeled turbulent forcing. Evaluating this modified operator requires determining the spatial distribution of $\nu_t$ in practice.

\begin{figure}
    \centering
    \def\svgwidth{0.8\textwidth}
    \fontsize{10}{11}
\begingroup%
  \makeatletter%
  \providecommand\color[2][]{%
    \errmessage{(Inkscape) Color is used for the text in Inkscape, but the package 'color.sty' is not loaded}%
    \renewcommand\color[2][]{}%
  }%
  \providecommand\transparent[1]{%
    \errmessage{(Inkscape) Transparency is used (non-zero) for the text in Inkscape, but the package 'transparent.sty' is not loaded}%
    \renewcommand\transparent[1]{}%
  }%
  \providecommand\rotatebox[2]{#2}%
  \newcommand*\fsize{\dimexpr\f@size pt\relax}%
  \newcommand*\lineheight[1]{\fontsize{\fsize}{#1\fsize}\selectfont}%
  \ifx\svgwidth\undefined%
    \setlength{\unitlength}{1064.4277882bp}%
    \ifx\svgscale\undefined%
      \relax%
    \else%
      \setlength{\unitlength}{\unitlength * \real{\svgscale}}%
    \fi%
  \else%
    \setlength{\unitlength}{\svgwidth}%
  \fi%
  \global\let\svgwidth\undefined%
  \global\let\svgscale\undefined%
  \makeatother%
  \begin{picture}(1,0.16125126)%
    \lineheight{1}%
    \setlength\tabcolsep{0pt}%
    \put(0,0){\includegraphics[width=\unitlength,page=1]{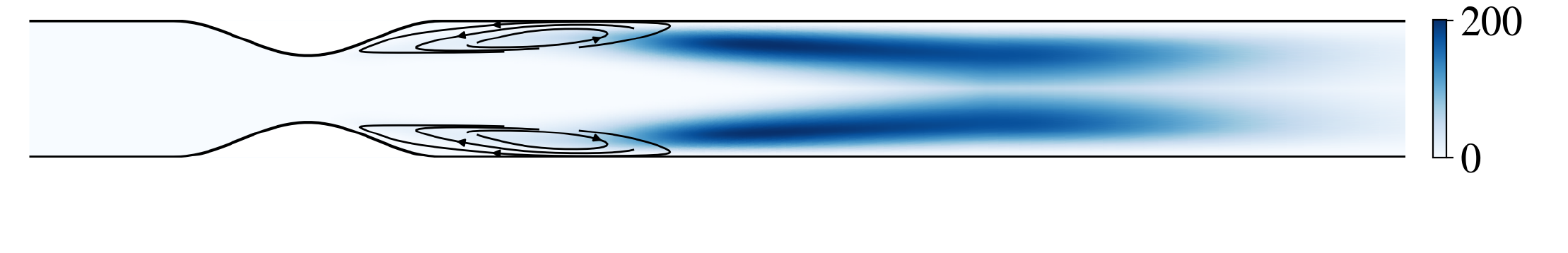}}%
    \put(0.97599964,0.09521592){\color[rgb]{0,0,0}\makebox(0,0)[lt]{\lineheight{0}\smash{\begin{tabular}[t]{l}$\nu_t/\nu$\end{tabular}}}}%
    \put(0.46990418,0.00373454){\color[rgb]{0,0,0}\makebox(0,0)[lt]{\lineheight{0}\smash{\begin{tabular}[t]{l}$x/d_t$\end{tabular}}}}%
    \put(0,0){\includegraphics[width=\unitlength,page=2]{nut.pdf}}%
  \end{picture}%
\endgroup%

    \caption{RANS-assimilated eddy viscosity field, together with the recirculation bubble streamlines.}
    \label{fig:nut}
\end{figure}

In practice, the eddy viscosity field is not accessible from a LES mean flow. Therefore, we assimilate a RANS solution to the LES mean field, using the methodology detailed and validated for the same flow in \citet{villie2025physics}.
A physics-informed neural network (PINN) infers this $\nu_t$ field from the LES mean flow by assimilating the Reynolds-averaged Navier-Stokes (RANS) equations, a method that has proven effective for data assimilation in separated flows \cite{von_saldern_mean_2022, patel2024turbulence, klopsch2025enabling}.
The present PINN formulation combines the axisymmetric RANS equations closed with the Spalart–Allmaras turbulence model.
Because the PINN-assimilated velocity field is virtually identical to the LES mean field, we chose to use only the assimilated eddy viscosity while retaining the LES mean velocity. 
Figure~\ref{fig:nut} presents the RANS-assimilated eddy viscosity field. 
Its magnitude is mainly located in the wake of the bubble. It remains low in the transitional domain and peaks in the turbulent breakdown region.

\subsection{Global linear stability analysis}\label{sec:LSA}
We first perform a global linear stability analysis (LSA) to characterize the asymptotic behavior of the unforced system, thereby revealing the intrinsic dynamical properties of the flow. 

Assuming the mean flow to be axisymmetric, a modal ansatz harmonic in $\theta$ and $t$ is chosen for the fluctuating quantities,
\begin{equation}\label{eq:ansatz}
    \bfq'(x,r,\theta,t) = \hbfq(x,r) \, \mathrm{e}^{i(m\theta - \omega t)} + \textrm{c.c.}, 
\end{equation}
where $\hbfq$ is the complex spatial mode shape, $i$ is the imaginary unit, $m \in \mathbb{Z}$ is the azimuthal wavenumber, $\omega \in \mathbb{C}$ is the angular frequency and c.c. is the complex conjugate.

The modified perturbation equations (\ref{eq:NS_perturbation1_mod}–\ref{eq:NS_perturbation2}) reduce to the homogeneous linear system
\begin{equation}\label{eq:GEVP}
-i\omega\, \boldsymbol{\mathsf{B}}
\left(\begin{array}{@{}c@{}}\hbfu\\ 
\hat{p}\end{array}\right) = \boldsymbol{\mathsf{L}}\,\left(\begin{array}{@{}c@{}}\hbfu\\ 
\hat{p}\end{array}\right),
\end{equation}
where $\boldsymbol{\mathsf{B}}$ is the mass matrix arising from the spatial discretization, and $\boldsymbol{\mathsf{L}}$ is the 2D linearized Navier–Stokes operator. $\boldsymbol{\mathsf{B}}$ and $\boldsymbol{\mathsf{L}}$ are detailed in Appendix~\ref{app:operators}.
Solving the generalized eigenvalue problem (\ref{eq:GEVP}) yields global modes and their complex eigenvalues $\omega = \omega_r + i\omega_i$, where $\omega_r$ denotes the oscillation frequency and $\omega_i$ the temporal growth rate.
Modes are classified as unstable if $\omega_i > 0$ or stable if $\omega_i < 0$. 

The stability analysis is performed using the in-house software FELiCS~\cite{kaiser_felics_2023}, a finite-element solver based on FEniCSx~\cite{alnaes2013fenics}, over a 2D structured grid of $8.2\times10^4$ triangular elements. 
At the two ends of the domain ($x/d_t< -4$ and $x/d_t > 6$), a quadratic sponge region damps all fluctuations through a dissipative term in the momentum equations, avoiding spurious reflections at the boundaries. We thus assume zero Dirichlet boundary conditions for all quantities at both the inlet and outlet of the mesh. The axis and wall are treated with axisymmetric and no-slip boundary conditions, respectively. 
FELiCS computes the eigenvalues and eigenvectors with the SLEPc library~\cite{hernandez2005slepc} and uses the Arnoldi \textit{shift-and-invert} algorithm to return the n-th closest eigenvalues to a prescribed initial guess.

\begin{figure*}%
    \centering
    \includegraphics[width=0.8\linewidth]{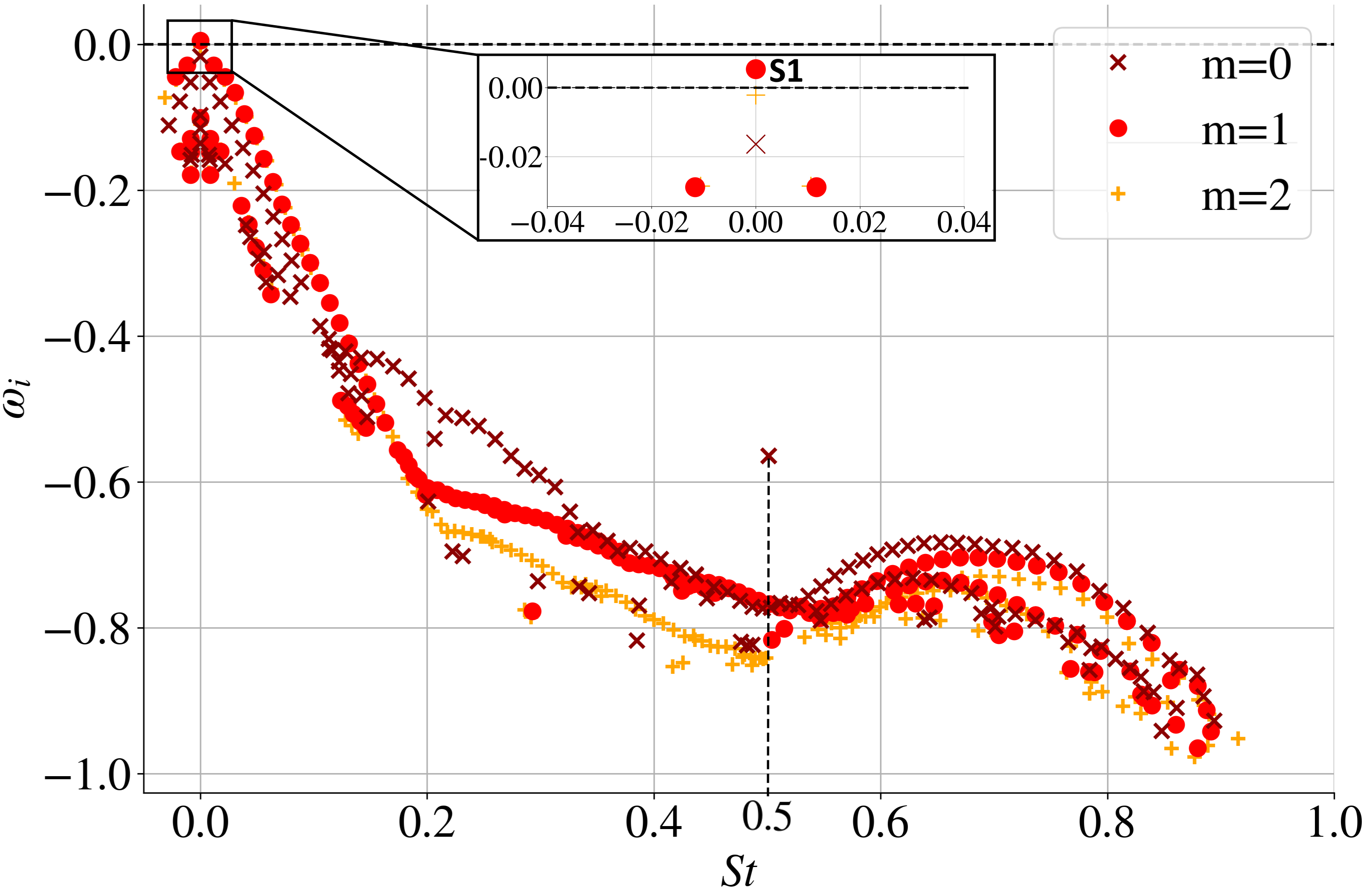}
    \caption{Global mode spectra for wavenumbers $m=0, 1, 2$. One positive growth rate is found at $m=1, \mathrm{St}=0$.}
    \label{fig:GLSA}
\end{figure*}

Figure~\ref{fig:GLSA} displays the eigenvalue spectra for azimuthal wavenumbers $m=0,1,2$ as a function of the Strouhal number $\mathrm{St} = \omega_r/2 \pi$.
Two main features emerge in the spectra. First, a discrete stationary mode lies near the stability axis for all $m$. Second, a distinct discrete mode appears at $\mathrm{St} = 0.50$ exclusively for $m = 0$ only. The rest of the spectrum is composed of stable continuous branches. 

\begin{figure*}
    \centering
    \def\svgwidth{0.7\textwidth}
\begingroup%
  \makeatletter%
  \providecommand\color[2][]{%
    \errmessage{(Inkscape) Color is used for the text in Inkscape, but the package 'color.sty' is not loaded}%
    \renewcommand\color[2][]{}%
  }%
  \providecommand\transparent[1]{%
    \errmessage{(Inkscape) Transparency is used (non-zero) for the text in Inkscape, but the package 'transparent.sty' is not loaded}%
    \renewcommand\transparent[1]{}%
  }%
  \providecommand\rotatebox[2]{#2}%
  \newcommand*\fsize{\dimexpr\f@size pt\relax}%
  \newcommand*\lineheight[1]{\fontsize{\fsize}{#1\fsize}\selectfont}%
  \ifx\svgwidth\undefined%
    \setlength{\unitlength}{1895.71046844bp}%
    \ifx\svgscale\undefined%
      \relax%
    \else%
      \setlength{\unitlength}{\unitlength * \real{\svgscale}}%
    \fi%
  \else%
    \setlength{\unitlength}{\svgwidth}%
  \fi%
  \global\let\svgwidth\undefined%
  \global\let\svgscale\undefined%
  \makeatother%
  \begin{picture}(1,0.21152621)%
    \lineheight{1}%
    \setlength\tabcolsep{0pt}%
    \put(0,0){\includegraphics[width=\unitlength,page=1]{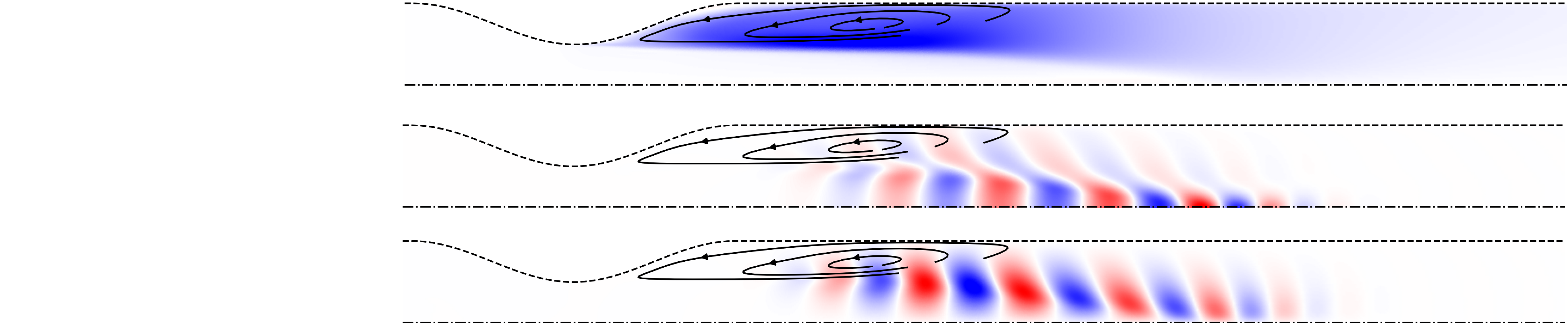}}%
    \put(0.04038006,0.18209754){\color[rgb]{0,0,0}\makebox(0,0)[lt]{\lineheight{0}\smash{\begin{tabular}[t]{l}Real$(\hbfu_x^{S1})$\end{tabular}}}}%
    \put(-0.00021953,0.1821901){\color[rgb]{0,0,0}\makebox(0,0)[lt]{\lineheight{0}\smash{\begin{tabular}[t]{l}a)\end{tabular}}}}%
    \put(-0.0005207,0.10165281){\color[rgb]{0,0,0}\makebox(0,0)[lt]{\lineheight{0}\smash{\begin{tabular}[t]{l}b)\end{tabular}}}}%
    \put(0.04038006,0.10255301){\color[rgb]{0,0,0}\makebox(0,0)[lt]{\lineheight{0}\smash{\begin{tabular}[t]{l}Real$(\hbfu_x^{KH})$\end{tabular}}}}%
    \put(0.04038006,0.03071538){\color[rgb]{0,0,0}\makebox(0,0)[lt]{\lineheight{0}\smash{\begin{tabular}[t]{l}Real$(\hbfu_r^{KH})$\end{tabular}}}}%
  \end{picture}%
\endgroup%

    \caption{Velocity components of the first stationary mode S1 at $m=1, \mathrm{St} = 0$ (a) and shear-layer mode at $m=0, \mathrm{St} = 0.5$ (b). Colour levels are normalised by the maximum amplitude in the domain. 
    }
    \label{fig:St0_mode_shape}
\end{figure*}
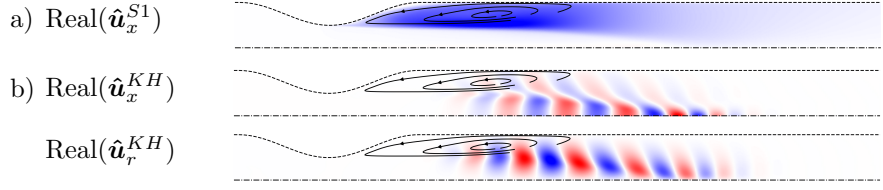

For all the wavenumbers considered, LSA reveals that the least damped mode is stationary ($\mathrm{St}=0$). Specifically, one stationary mode, denoted S1, is amplified ($\omega_i >0$) for $m=1$. The axial velocity component of this leading stationary mode, shown in figure~\ref{fig:St0_mode_shape}(a), is localized primarily in the shear-layer and inside the separation bubble. 
The axial component of this $m=1$ stationary perturbation accelerates the flow at one side of the stenosis, and decelerates it at the other side, thus deflecting the confined jet away from the stenosis axis. 
In their LSA of the same flow at $Re = 750$, \citet{sherwin2005three} identified stationary eigenmodes for $m\geq1$ with similar structure, finding $m=1$ to be unstable. 
They describe it as a mild Coanda-type instability breaking the axisymmetry of the flow, in which the jet attaches to one side of the stenosis. The same eigenmode was also found in the 3D LSA of the same geometry at $Re=721$ \cite{loiseau2014dynamics}. 
Such a mild Coanda-type instability seems to persist beyond transition, as observed here at $Re=4000$. 

Since then, stationary eigenmodes with comparable features have been documented in several separated flows with recirculation bubbles, including stall cells \cite{sarras2024linear}, pressure-induced separation bubbles \cite{wu2020spatio, cura2024low, fuchs2025standing} and separated flows over a gaussian bump \cite{klopsch2025enabling}. Recent studies have linked the stationary mode to a centrifugal instability \cite{Savarino_Sipp_Rigas_2025, fuchs2025standing}, a mechanism known to accelerate transition in laminar separation bubbles \cite{gallaire2007three, rodriguez2013two}. Although these observations only concern planar configurations, we hypothesize, based on the structural similarity with the S1 mode, that a similar underlying mechanism may be at play in the present axisymmetric case.

Since stability analysis is performed on a mean flow, the growth rate of the S1 mode carries inherent uncertainty and warrants cautious interpretation. Some inconsistencies in the equation closure might come into play. Modeling inconsistencies, such as the frozen eddy viscosity assumption discussed in \citet{sarras2024linear}, may affect its exact value, and we hypothesize that a fully consistent model would lead to a marginally stable S1 mode.
Nevertheless, this stationary mode remains of interest, as it influences the low-frequency response of the flow as demonstrated in section~\ref{sec:resolvent}.

The LSA eigenspectrum in figure~\ref{fig:GLSA} identifies one discrete mode at $m=0, \mathrm{St} = 0.50$. 
While stable, this mode might have an influence on the dynamics and be sensitive to forcing.
Its structure is a compact wavepacket that grows in the shear-layer and gradually decays downstream of the recirculation bubble [see figure~\ref{fig:St0_mode_shape}(b)].
Its phase velocity approximately $U_t/2$ in the center of the shear-layer, which is characteristic for Kelvin–Helmholtz instability modes \citep{suzuki2006instability}. 

Because the Kelvin–Helmholtz mechanism drives a convective instability, one expects it to manifest in the global eigenvalue spectrum as a continuous branch of shear-layer modes, as observed in unconfined jets \citep{garnaud2013preferred}.
However, multiple studies report global discrete eigenmodes near $St \approx 0.5$ in various flows featuring a free shear-layer. For instance, \citet{loiseau2014dynamics} identified comparable discrete eigenvalues around $St \approx 0.4$ in a 3D LSA of the same stenotic flow at $Re = 721$. Furthermore, global LSA of coaxial jets with backflow reveals an $m=0$ eigenmode within the shear-layer that exhibits a similar structure near the stability threshold \citep{canton2017linear}. Additionally, the separated shear-layer above an airfoil exhibits a dominant eigenmode near $St \approx 0.7$ \cite{yeh2019resolvent}.
Even damped, finding such a discrete shear-layer eigenmode may indicate a preferred frequency for the vortex shedding.

The global LSA reveals an unstable stationary mode arising from a mild Coanda-type instability and a stable, discrete shear-layer mode. 

\subsection{Resolvent analysis}\label{sec:resolvent}
Although LSA reveals the intrinsic dynamics of the flow, it cannot capture the transient growth mechanisms excited by nonlinear turbulent forcing. Because this non-modal growth significantly influences stenotic flow dynamics \citep{griffith2008steady}, we employ resolvent analysis (RA) to identify these amplification mechanisms. RA characterizes the input-output response of the linearized system to harmonic forcing within a time-invariant flow.

Starting from the perturbation equations~(\ref{eq:NS_perturbation1}–\ref{eq:NS_perturbation2}) and using the same temporal–azimuthal modal ansatz (Eq.~\ref{eq:ansatz}), the system in the frequency domain becomes
\begin{equation}
\hbfu = \mathcal{R}\, \hbff,
\end{equation}
introducing the resolvent operator
\begin{equation}\label{eq:RA}
\mathcal{R} = \boldsymbol{\mathsf{D}}_r\boldsymbol{\mathsf{P}}^T \,( -i\omega \boldsymbol{\mathsf{B}} - \boldsymbol{\mathsf{L}} )^{-1} \boldsymbol{\mathsf{P}}\boldsymbol{\mathsf{D}}_f,
\end{equation}
where 
$\boldsymbol{\mathsf{P}}$ removes the forcing in the mass equation and the pressure component (detailed in Appendix~\ref{app:operators}). 
The $\boldsymbol{\mathsf{D}}_f$ and $\boldsymbol{\mathsf{D}}_r$ operators allow one to select the spatial regions considered in the definition of the output and input, respectively.
Both regions are confined to the immediate post-stenotic region $x/d_t \in [-2, 4]$, before jet reattachment and vortical breakdown. 

A singular value decomposition of the resolvent operator $\mathcal{R}$ at a given $\left(m, \omega\right)$ yields
\begin{equation}\label{eq:SVD}
\mathcal{R} = \sum_j \sigma^{(j)} \, \hbfu^{(j)} \, \hbff^{(j)^*},
\end{equation}
where the right and left singular vectors, $\hbff^{(j)}$ and $\hbfu^{(j)}$, represent the $j$-th optimal forcing and response modes, respectively, and the singular values $\sigma^{(j)}$ quantify the energetic gain of each corresponding pair.
The problem is solved with the same numerical method, mesh and boundary conditions as the LSA.

RA is rigorously applicable to asymptotically stable flows where the linearized Navier-Stokes operator exhibits eigenvalues with $\omega_i < 0$. 
Therefore, the globally unstable stationary mode (S1) identified in the LSA [see figure~\ref{fig:GLSA}] is stabilized by tuning the eddy viscosity. We chose to introduce a multiplicative factor $\gamma$ to the PINN-assimilated eddy viscosity field:
\begin{equation}
    \nu_{t}(x, r) = \gamma \, \nu_{t, PINN}(x, r).
\end{equation}
We set $\gamma = 0.5$, which corresponds to the minimal modification of $\nu_{t, PINN}(x, r)$ that stabilizes the S1 mode, enabling RA.
%

\begin{figure*}%
    \centering
    \def\svgwidth{1\textwidth}
    \fontsize{12}{1}\selectfont
    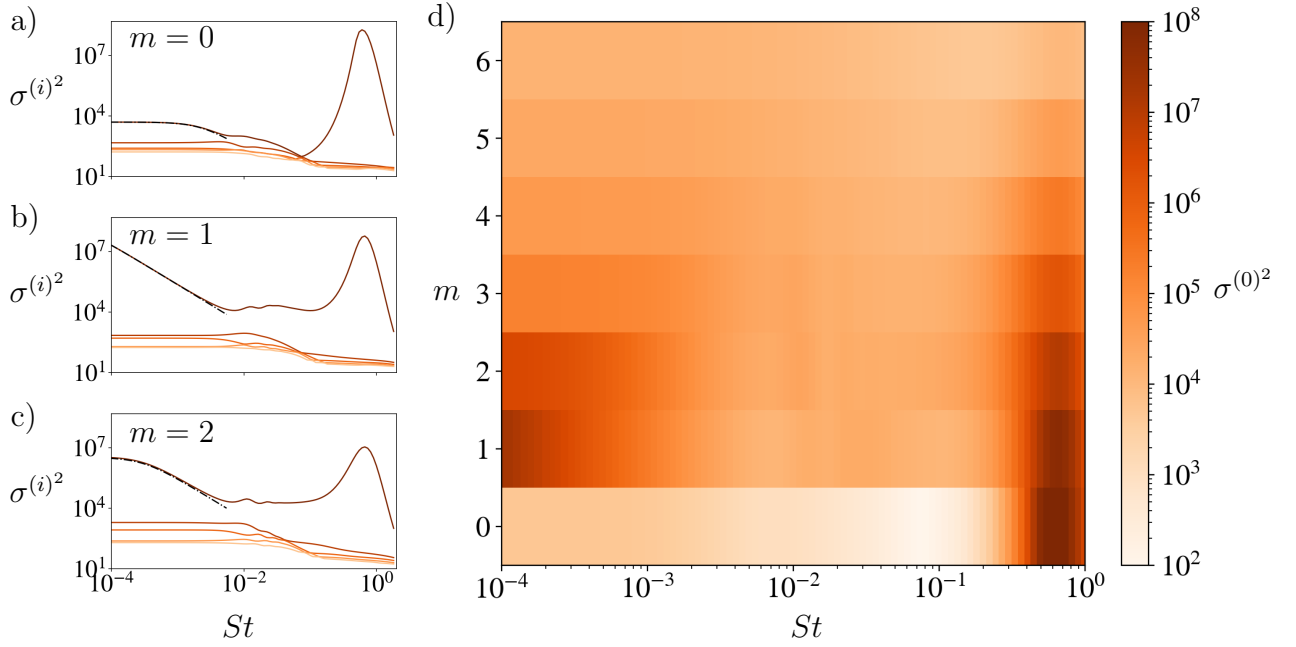
    \caption{Resolvent gain spectra for the first five resolvent modes for $m = 0,1,2$ (a-c), the low-pass filter model from \citet{bugeat2022low} is added in dashed lines. Heatmap of the optimal resolvent gain as a function of $(m, \, \mathrm{St})$ (d).}
    \label{fig:St_m_RA_gains}
\end{figure*}

Figure~\ref{fig:St_m_RA_gains}(a-c) shows the gains of the first five resolvent modes as a function of $\mathrm{St}$ for $m = 0,1,2$, while figure~\ref{fig:St_m_RA_gains}(d) highlights the regions of strongest amplification by plotting the optimal gain in the $(m,\mathrm{St})$ plane.
\begin{figure}
  \centering 
  \def\svgwidth{1\textwidth}
  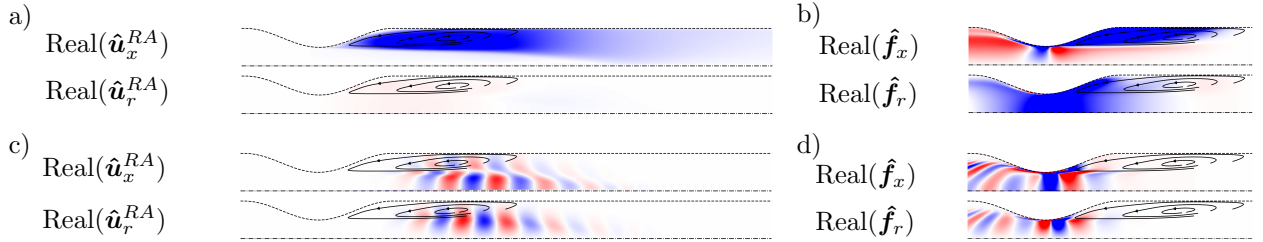
  \caption{Mode shapes of the optimal resolvent forcing and response at $m=1, \, \mathrm{St}=0.0001$ (a, b) and $m=0, \, \mathrm{St}=0.48$ (c, d). The panels show the real part of the axial and radial velocity components, together with the recirculation bubble streamlines. 
  Response mode colour levels scale with their respective maxima, whereas forcing mode colour levels saturate at one-hundredth of their maxima to emphasize their upstream spread.
  }
  \label{fig:Mode_shapes_RA}
\end{figure}
A first region of significant amplification is identified in the low frequency regime.  Figure~\ref{fig:St_m_RA_gains}(a-c) show a high gain separation for $m=0, 1,$ and $2$ at $\mathrm{St} < 0.01$. Figure~\ref{fig:St_m_RA_gains}(d) demonstrates
that this amplification is more pronounced for $m=1$ but seems to occur for all $m \geq 1$. 
Similar low-pass filter behavior has been demonstrated to emerge from the stationary mode S1 in several studies considering separation bubbles \cite{robinet2007bifurcations, theofilis2000origins, cura2024low}. 
The closer the stationary mode is to the stability axis, the lower the cutoff frequency of the low-pass filter. 
The influence of the S1 mode on the resolvent gains has been modeled by \citet{bugeat2022low}. The low-frequency amplification caused by the stationary eigenvalue at first order is:
\begin{equation} 
    \sigma (\mathrm{St}) = \frac{\sigma^{(0)}(\mathrm{St}\rightarrow 0)}{\sqrt{1+ \left( \dfrac{\mathrm{St}}{\omega_i^{(S1)}/2 {\rm \pi}} \right)^2}}, 
\end{equation}
where $\sigma^{(0)}(\mathrm{St}\rightarrow 0)$ is the value of the optimal gain when $\mathrm{St}$ approaches zero, and $\omega_i^{(S1)}$ is the growth rate of the S1 mode.
This model is plotted in dashed lines in figure~\ref{fig:St_m_RA_gains}(a-c) and closely matches the low frequency gains for the first three azimuthal wavenumbers. 
The optimal response 
mode shape at $m=1, \, \mathrm{St}=0.0001$ is shown in figure~\ref{fig:Mode_shapes_RA}(a).
It covers the recirculation bubble and is qualitatively similar to the stationary S1 mode.
This confirms that low-frequency amplification arises from the forced response to the S1 mode. 
Consequently, we expect the confined jet to exhibit a meandering motion at frequencies below $\mathrm{St} = 0.01$.
%

A second region of significant amplification is observed across $\mathrm{St} \in [0.1, 1]$, where the optimal gain exhibits high separation to the suboptimal gains. This prevalent broadband amplification reflects the convective instability of the separated shear-layer, with maximum amplification occurring at $m=0$, $\mathrm{St} = 0.59$, and decreasing monotonically as $m$ increases. The structure of the $m=0$ optimal response mode, shown in figure~\ref{fig:Mode_shapes_RA}(c), closely matches the shear-layer eigenmode detected in the LSA. The forcing shown in figure~\ref{fig:Mode_shapes_RA}(d) is located close to the separation point, and drives a downstream response that amplifies in the separated shear-layer, reaching maximum amplitude at $x/d_t = 3.5$, corresponding to the center of the recirculation bubble. In our LES snapshots, this manifests as axisymmetric 2D rolled-up vortices developing in the shear-layer at $\mathrm{St} = 0.5$ [see figure~\ref{fig:Q_c_mean_flow}(a)].

This broadband behavior is typical of amplifier flows \cite{alizard2009sensitivity, sipp2013characterization, beneddine_conditions_2016, garnaud2013preferred, thomareis2018resolvent}, and the forcing and response structure is reminiscent of the convective Kelvin--Helmholtz instability in turbulent jets at similar $(m, \, \mathrm{St})$ \citep{garnaud2013preferred, pickering2020lift}. The dominance of $m=0$ fluctuations contrasts with the transient growth analysis of \citet{blackburn2008convective} at $Re = 750$, 
who identified sinuous waves (i.e. $m=1$) as the most amplified mode, as confirmed experimentally by \citet{griffith2010convective}.
However, it agrees with the axisymmetric rollers reported at $Re=1000$ in both experiments \citep{vetel2008asymmetry} and DNS \citep{varghese2007direct}.
%

The resolvent analysis reveals two distinct amplification mechanisms. At low frequencies ($\mathrm{St} < 0.01$), the flow acts as a low-pass filter driven by the S1 eigenmode. At intermediate frequencies ($\mathrm{St} \in [0.1, 1]$), broadband convective amplification dominates, driven by the Kelvin-Helmholtz instability of the separated shear-layer, and peaks at $m=0$, $\mathrm{St} = 0.59$. 

\subsection{Pseudospectral analysis}\label{sec:pseudospectrum}
An observation that deserves further investigation is the apparent proximity of the frequency and structure between the RA broadband amplification and the discrete eigenvalue identified in the LSA at $m=0, \, \mathrm{St} = 0.5$.
We thus raise the question: 
Is the broadband amplification in the resolvent gains caused only by the shear-layer eigenmode (modal mechanism), or is it the result of a pseudoresonance arising from the non-normal interaction of a combination of modes?
This section investigates the influence of the discrete shear-layer eigenmode on the resolvent gain. For this, we compute the pseudospectrum, which characterizes the non-normality of the eigenmodes.

The pseudospectrum is mapped by evaluating the optimal resolvent gain over the complex frequency plane $\omega = \omega_r + \mathrm{i}\omega_i$.
Computing $\sigma(\omega_r, \, \omega_i)$ across a grid of complex frequencies directly yields the pseudospectrum. In this formulation, the imaginary part $\omega_i$ acts as a discounting parameter applied to the forcing and response, imposing a finite-time horizon for the evaluation of the input-output gain. Extracting slices at a constant $\omega_i$ is therefore mathematically equivalent to performing the discounted resolvent analysis introduced by \citet{Jovanovic2004}.

\begin{figure}
    \centering
    \includegraphics[width=0.7\linewidth]{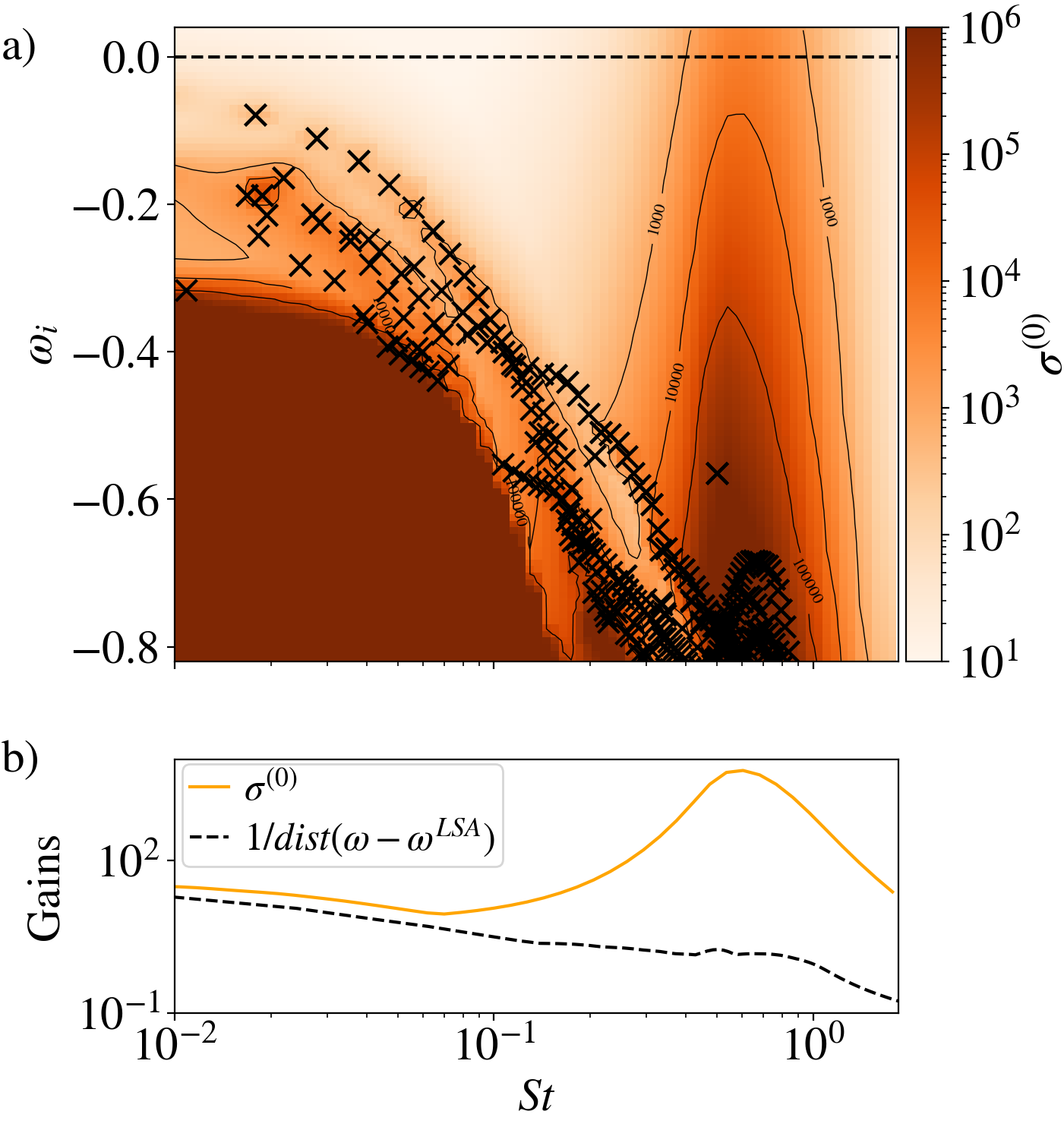}
    \caption{Pseudospectrum for $m=0$ along with the least damped eigenvalues (a). Resolvent 
    optimal gains and inverse distance from the imaginary axis to the nearest eigenvalue (dashed line) (b). The resolvent norm reflects the value of the pseudospectrum along the real axis.}
    \label{fig:pseudospectrum}
\end{figure}

Figure~\ref{fig:pseudospectrum}(a) shows the pseudospectrum of the resolvent operator at $m=0$ along with the LSA eigenvalues. 
The pseudospectrum identifies that the broadband amplification arises not only from the shear-layer eigenmode but also from the collective contribution of several more strongly damped eigenvalues.
Figure~\ref{fig:pseudospectrum}(b) presents the optimal resolvent gain, 
along with the inverse distance from the real axis to the closest eigenvalue in dashed line. 
It shows the amplification gain that a normal operator would exhibit, following \cite{symon2018non}. Close correspondence of both lines highlight that the amplification is dominated by the normal response of one eigenmode.   
This hypothetical "normal" behavior presents a small peak at $\mathrm{St} = 0.5$, showing how a normal operator would resonate with the shear-layer eigenvalue. The substantial gap with the actual resolvent gain demonstrates that the pseudoresonance dominates over the resonance of the shear-layer eigenmode.
Consequently, this confirms that the broadband amplification in the resolvent gains is not an isolated modal resonance of the shear-layer eigenmode, but rather arise from strong pseudoresonance driven by the highly non-normal interaction of multiple damped modes.

\section{Data-driven analysis of coherent structures}\label{sec:data_driven} 

This section aims to validate the linear model with a SPOD of the LES snapshots. 
We employ SPOD as introduced in~\citet{towne_spectral_2018} to isolate the most energetic time-azimuthal-coherent structures.
 
The 3D velocity fluctuations $\bfu'$ are computed from the LES snapshots and decomposed into azimuthal Fourier modes: 
\begin{equation}
    \hbfu(x, r, m, t) = \int_{-\pi}^{\pi} \bfu'(x, r, \theta, t) \, e^{-im \theta} \, \text{d}\theta.
\end{equation}
The azimuthal coefficients are Fourier transformed into the frequency domain, and a basis is defined by the eigenvalue decomposition of the cross spectral density matrix. 
Due to the high spatial resolution of the computational grid, the snapshots method~\cite{schmidt_guide_2020} is used from an inner product of two realizations $\hbfu, \hbfv$: 
\begin{equation}
	\langle \hbfu, \hbfv\rangle = \int_{\Omega}\hbfu^*\hbfv d\mathbf{V}= \hbfu^*\textbf{W}\hbfv\text{,}
    \label{eq:inner_product}
\end{equation}
where $\Omega$ is the region of interest, $\textbf{W}$ is the discretized weighting operator, and $\cdot^*$ denotes the conjugate transpose. We chose a weighted 2-norm 
to obtain a measure of the turbulent kinetic energy \cite{schmidt_guide_2020}. 
The following analyses are performed with the \textsc{PySPOD} code \citep{rogowski2024unlocking}, using $29$ blocks with an overlap of $75\%$, resulting in $512$ frequency bins per block and a frequency resolution $\Delta \mathrm{St} = 0.02$.

The discretized weighting operator is modified to enable extended SPOD, as done for POD in \citet{boree2003extended}. By setting selected entries of $\textbf{W}$ to $0$, the energy norm used in the SPOD formulation can be confined to a targeted spatial subdomain. In this manner, the energy calculated by the inner product is only evaluated in the region of interest, while the SPOD modes are reconstructed over the whole domain as linear combinations of flow realizations. This approach is used to capture the coherent structures impacting only a certain region.
%

In our stenotic flow, the initial growth of coherent vortical structures and their turbulent breakdown occupy two distinct regions. 
To separately analyze the transitional and turbulent dynamics, we define two streamwise subdomains: a transitional domain at $x/d_t \in [-2, 4]$ and a turbulent domain at $x/d_t \in [5, 12]$.
Extended SPOD is performed separately in each region by restricting the discretized weighting operator, $\textbf{W}$, to the corresponding subdomain.

\begin{figure}
  \centering 
  \def\svgwidth{1\textwidth}
  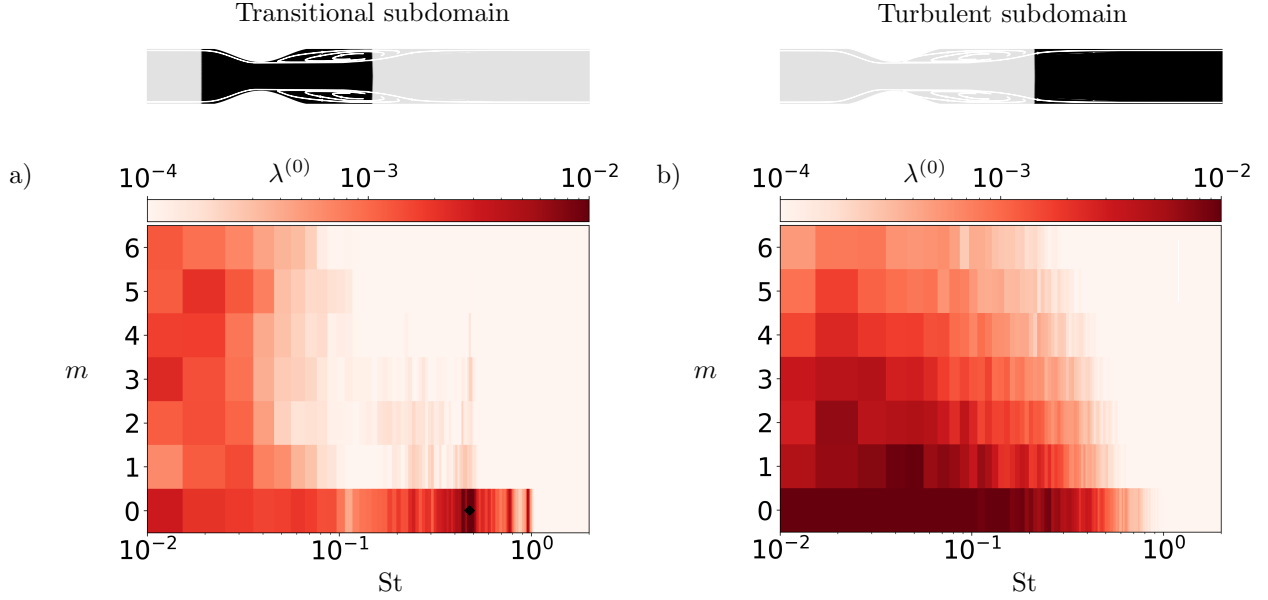
  \caption{Color map of the leading SPOD eigenvalues for the first azimuthal wavenumbers. The two extended SPOD present the transitional (a) and turbulent (b) flow dynamics by selecting the weights in the black-colored regions. The marker $\blacklozenge$ denotes the highest energy area at $m=0, \, \mathrm{St}=0.48$ in the transitional domain.}
  \label{fig:spod_spectrum_colormap}
\end{figure}
\begin{figure}
  \centering 
  \def\svgwidth{1\textwidth}
  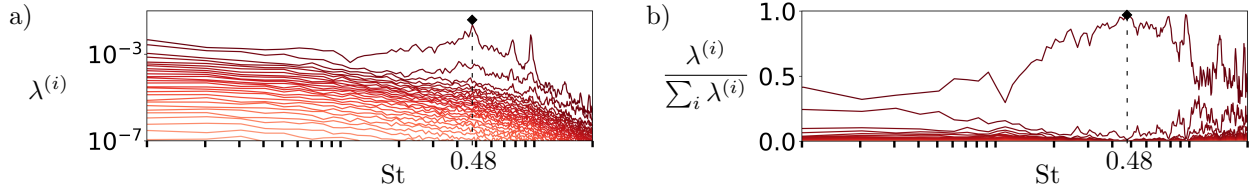
  \caption{SPOD eigenvalues (a) and relative eigenvalues (b) for $m=0$ in the transitional subdomain. The marker $\blacklozenge$ denotes the highest energy area at the shedding frequency $\mathrm{St}=0.48$.} 
  \label{fig:spod_spectrum_m0}
\end{figure}

The color levels of the leading SPOD eigenvalues shown in figure~\ref{fig:spod_spectrum_colormap} illustrate the flow dynamics across the transitional and turbulent subdomains as a function of $m, \, \mathrm{St}$.
The turbulent subdomain distributes energy over a broad range of frequencies and azimuthal wavenumbers. This broadband energy spectrum indicates that the vortices break into turbulence, leaving no dominant coherent structure in this region. In contrast, the transitional subdomain displays a concentration of TKE for $m=0$ and mainly in the range $\mathrm{St} \in [0.1, 1]$. A strong TKE peak is observed at $\mathrm{St}=0.48$, which corresponds to the frequency of the vortex rings identified in figure~\ref{fig:Q_c_mean_flow}.
As $m$ increases, the energy peaks decrease significantly in amplitude. 
Consequently, the SPOD optimized in the transitional subdomain provides a more suitable basis for representing the vortices dynamics, and all subsequent SPOD results refer to this decomposition.

Focusing on the $m=0$ component in the transitional subdomain, figure~\ref{fig:spod_spectrum_m0} plots the SPOD spectrum and relative energy distribution. Two features emerge. 
First, a broadband amplification spans $\mathrm{St} \in [0.1, 1]$. On average over this range, the leading SPOD mode accounts for 82\% of the total TKE, demonstrating substantial gain separation from the sub-leading mode. 
Second, the spectrum displays sharp peaks at the shedding frequency $\mathrm{St}=0.48$ and its harmonics ($\mathrm{St}=0.96$ and $1.92$), along with half of the shedding frequency and its harmonics ($\mathrm{St}=0.24$ and $0.78$). The relative energy distribution further emphasizes this dominance, showing that the highest peak at $\mathrm{St}=0.48$ accounts for $98\%$ of the total turbulent kinetic energy at this frequency. These sharp peaks indicate a resonance mechanism typical of oscillator-type flows. Experimental studies \citep{cassanova1978disorder, vetel2008asymmetry} and direct numerical simulations \citep{varghese2007direct} report similar spectral peaks at $\mathrm{St}=0.5$ in $75\%$ stenosis flows for $Re \sim 1000$, which represent the signature of passing vortex rings.

\begin{figure}
  \centering 
  \def\svgwidth{0.6\textwidth}
\begingroup%
  \makeatletter%
  \providecommand\color[2][]{%
    \errmessage{(Inkscape) Color is used for the text in Inkscape, but the package 'color.sty' is not loaded}%
    \renewcommand\color[2][]{}%
  }%
  \providecommand\transparent[1]{%
    \errmessage{(Inkscape) Transparency is used (non-zero) for the text in Inkscape, but the package 'transparent.sty' is not loaded}%
    \renewcommand\transparent[1]{}%
  }%
  \providecommand\rotatebox[2]{#2}%
  \newcommand*\fsize{\dimexpr\f@size pt\relax}%
  \newcommand*\lineheight[1]{\fontsize{\fsize}{#1\fsize}\selectfont}%
  \ifx\svgwidth\undefined%
    \setlength{\unitlength}{358.27685499bp}%
    \ifx\svgscale\undefined%
      \relax%
    \else%
      \setlength{\unitlength}{\unitlength * \real{\svgscale}}%
    \fi%
  \else%
    \setlength{\unitlength}{\svgwidth}%
  \fi%
  \global\let\svgwidth\undefined%
  \global\let\svgscale\undefined%
  \makeatother%
  \begin{picture}(1,0.82149272)%
    \lineheight{1}%
    \setlength\tabcolsep{0pt}%
    \put(0,0){\includegraphics[width=\unitlength,page=1]{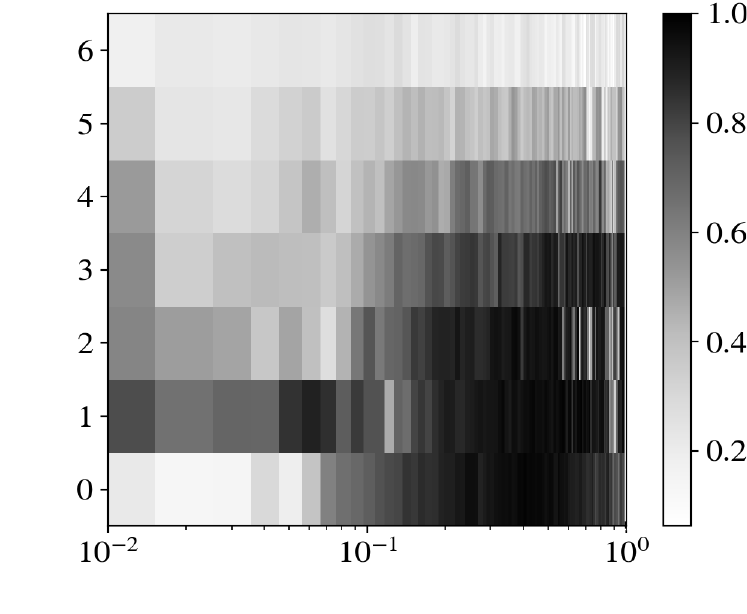}}%
    \put(0.00199342,0.44964194){\color[rgb]{0,0,0}\makebox(0,0)[lt]{\lineheight{0}\smash{\begin{tabular}[t]{l}$m$\end{tabular}}}}%
    \put(0.48795023,0.01229715){\color[rgb]{0,0,0}\makebox(0,0)[lt]{\lineheight{0}\smash{\begin{tabular}[t]{l}$St$\end{tabular}}}}%
  \end{picture}%
\endgroup%

  \caption{First SPOD and resolvent modes alignment.}
  \label{fig:alignment}
\end{figure}

A quantitative comparison between the velocity modes from the leading SPOD and the optimal RA modes is obtained by computing their alignment using the inner product in Eq.~\ref{eq:inner_product}:
\begin{align}
    Q(m, \omega) &= \frac{ |\langle \hbfu_\text{SPOD}(m, \omega), \hbfu_\text{RA}(m, \omega) \rangle |}{\| \hbfu_\text{SPOD}(m, \omega) \| \cdot \| \hbfu_\text{RA}(m, \omega)  \|}.
\end{align}
Figure~\ref{fig:alignment} reports this alignment over the $(m, \mathrm{St})$-space.
A strong alignment is found in the region of convective amplification in the range $\mathrm{St} \in [0.1,1]$ for $m\leq 3$, confirming that the dominant coherent structures are driven by linear mechanisms. 
A good alignment is also observed in the lower frequencies for $m=1$. However, the comparison between the SPOD and the linear analysis is limited by LES duration in the low frequency range. As only the frequencies $\mathrm{St} > 0.01$ are available in the SPOD, the comparison with the low frequency modal mechanism happening for $\mathrm{St} < 0.01$ is made impossible.

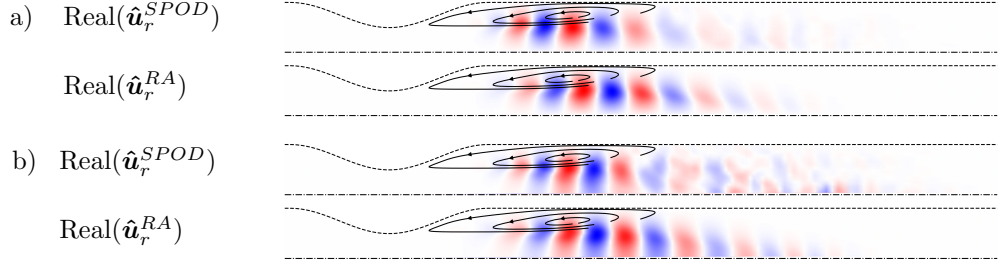
\begin{figure}
  \centering 
  \def\svgwidth{0.8\textwidth}
\begingroup%
  \makeatletter%
  \providecommand\color[2][]{%
    \errmessage{(Inkscape) Color is used for the text in Inkscape, but the package 'color.sty' is not loaded}%
    \renewcommand\color[2][]{}%
  }%
  \providecommand\transparent[1]{%
    \errmessage{(Inkscape) Transparency is used (non-zero) for the text in Inkscape, but the package 'transparent.sty' is not loaded}%
    \renewcommand\transparent[1]{}%
  }%
  \providecommand\rotatebox[2]{#2}%
  \newcommand*\fsize{\dimexpr\f@size pt\relax}%
  \newcommand*\lineheight[1]{\fontsize{\fsize}{#1\fsize}\selectfont}%
  \ifx\svgwidth\undefined%
    \setlength{\unitlength}{1965.7862866bp}%
    \ifx\svgscale\undefined%
      \relax%
    \else%
      \setlength{\unitlength}{\unitlength * \real{\svgscale}}%
    \fi%
  \else%
    \setlength{\unitlength}{\svgwidth}%
  \fi%
  \global\let\svgwidth\undefined%
  \global\let\svgscale\undefined%
  \makeatother%
  \begin{picture}(1,0.26209989)%
    \lineheight{1}%
    \setlength\tabcolsep{0pt}%
    \put(0.05342318,0.2372357){\color[rgb]{0,0,0}\makebox(0,0)[lt]{\lineheight{0}\smash{\begin{tabular}[t]{l}Real$(\hbfu_r^{SPOD})$\end{tabular}}}}%
    \put(0.05342318,0.16785863){\color[rgb]{0,0,0}\makebox(0,0)[lt]{\lineheight{0}\smash{\begin{tabular}[t]{l}Real$(\hbfu_r^{RA})$\end{tabular}}}}%
    \put(0.00170486,0.09202429){\color[rgb]{0,0,0}\makebox(0,0)[lt]{\lineheight{0}\smash{\begin{tabular}[t]{l}b)\end{tabular}}}}%
    \put(-0.0003405,0.2372357){\color[rgb]{0,0,0}\makebox(0,0)[lt]{\lineheight{0}\smash{\begin{tabular}[t]{l}a)\end{tabular}}}}%
    \put(0.04933371,0.09202429){\color[rgb]{0,0,0}\makebox(0,0)[lt]{\lineheight{0}\smash{\begin{tabular}[t]{l}Real$(\hbfu_r^{SPOD})$\end{tabular}}}}%
    \put(0.04933371,0.02264762){\color[rgb]{0,0,0}\makebox(0,0)[lt]{\lineheight{0}\smash{\begin{tabular}[t]{l}Real$(\hbfu_r^{RA})$\end{tabular}}}}%
    \put(0,0){\includegraphics[width=\unitlength,page=1]{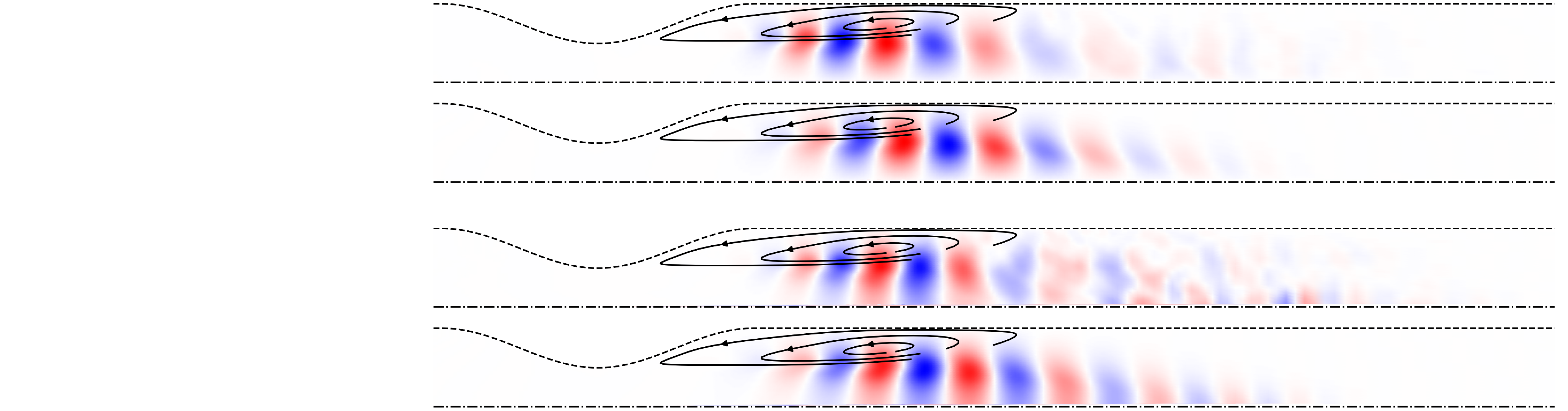}}%
  \end{picture}%
\endgroup%

  \caption{Leading SPOD and optimal resolvent mode shapes at $\mathrm{St}=0.48$ for $m=0$ (a) and $m=1$ (b). The panels show the real part of the radial velocity component, together with the recirculation bubble streamlines. The colormap is saturated at the maximum absolute amplitude.}
  \label{fig:spod_RA_modes}
\end{figure}

The SPOD and RA velocity mode shapes at the shedding frequency $\mathrm{St}=0.48$ are compared in figure~\ref{fig:spod_RA_modes} for $m=0,1$.
The leading SPOD mode and the optimal RA modes are in close agreement. 
All mode shapes examined in the frequency range $\mathrm{St}\in [0.1, 1]$ are confined in the shear-layer and exhibit similar spatial structures. Together with the strong gain-separation in this frequency range, this indicates that the Kelvin-Helmholtz instability of the shear-layer dominates the dynamics, in agreement with the turbulent jet analysis of \citet{pickering2020lift}.

\begin{figure}
  \centering 
  \def\svgwidth{0.7\textwidth}
\begingroup%
  \makeatletter%
  \providecommand\color[2][]{%
    \errmessage{(Inkscape) Color is used for the text in Inkscape, but the package 'color.sty' is not loaded}%
    \renewcommand\color[2][]{}%
  }%
  \providecommand\transparent[1]{%
    \errmessage{(Inkscape) Transparency is used (non-zero) for the text in Inkscape, but the package 'transparent.sty' is not loaded}%
    \renewcommand\transparent[1]{}%
  }%
  \providecommand\rotatebox[2]{#2}%
  \newcommand*\fsize{\dimexpr\f@size pt\relax}%
  \newcommand*\lineheight[1]{\fontsize{\fsize}{#1\fsize}\selectfont}%
  \ifx\svgwidth\undefined%
    \setlength{\unitlength}{946.21429203bp}%
    \ifx\svgscale\undefined%
      \relax%
    \else%
      \setlength{\unitlength}{\unitlength * \real{\svgscale}}%
    \fi%
  \else%
    \setlength{\unitlength}{\svgwidth}%
  \fi%
  \global\let\svgwidth\undefined%
  \global\let\svgscale\undefined%
  \makeatother%
  \begin{picture}(1,0.43854807)%
    \lineheight{1}%
    \setlength\tabcolsep{0pt}%
    \put(0,0){\includegraphics[width=\unitlength,page=1]{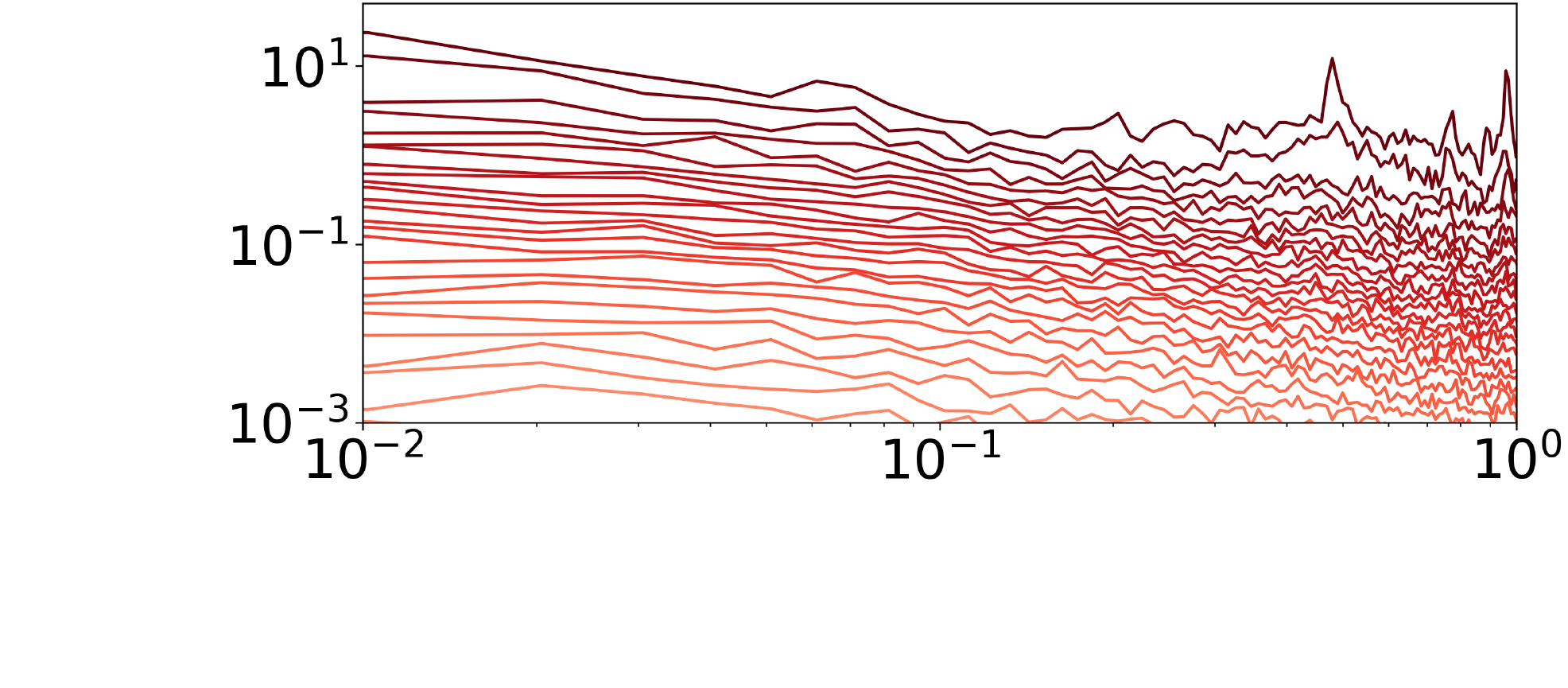}}%
    \put(0.58625602,0.09935645){\color[rgb]{0,0,0}\makebox(0,0)[lt]{\lineheight{0}\smash{\begin{tabular}[t]{l}$\mathrm{St}$\end{tabular}}}}%
    \put(-0.00020505,0.03153702){\color[rgb]{0,0,0}\makebox(0,0)[lt]{\lineheight{0}\smash{\begin{tabular}[t]{l}b)\end{tabular}}}}%
    \put(0.00039357,0.39364229){\color[rgb]{0,0,0}\makebox(0,0)[lt]{\lineheight{0}\smash{\begin{tabular}[t]{l}a)\end{tabular}}}}%
    \put(0.83117342,0.41256018){\color[rgb]{0,0,0}\makebox(0,0)[lt]{\lineheight{0}\smash{\begin{tabular}[t]{l}$0.48$\end{tabular}}}}%
    \put(0,0){\includegraphics[width=\unitlength,page=2]{forcing_SPOD_spectrum.pdf}}%
    \put(0.06322338,0.27715823){\color[rgb]{0,0,0}\makebox(0,0)[lt]{\lineheight{0}\smash{\begin{tabular}[t]{l}$\lambda^{(i)}$\end{tabular}}}}%
    \put(0.06225755,0.03133795){\color[rgb]{0,0,0}\makebox(0,0)[lt]{\lineheight{0}\smash{\begin{tabular}[t]{l}Real$(\hbff_x)$\end{tabular}}}}%
  \end{picture}%
\endgroup%

  \caption{SPOD spectrum of the forcing $\hbff = \hbfu \nabla \hbfu$ for $m=0$ fluctuations (a) and leading mode shape at $\mathrm{St}=0.48$ (b). The panels show the real part of the axial component, together with the recirculation bubble streamlines. The colormap is saturated at the maximum absolute amplitude.}
  \label{fig:forcing_SPOD_spectrum}
\end{figure}

The sharp peak observed in the SPOD spectrum at $m=0, \mathrm{St}=0.5$ and its harmonics shown in figure~\ref{fig:spod_spectrum_m0}(a) are missing in the resolvent spectrum shown in figure~\ref{fig:St_m_RA_gains}. 
We expect these peaks to be the result of a correlated forcing, characteristic in transitional flows \citep{symon_tale_2019}, in contrast to the resolvent model, which assumes the true forcing to be white noise in frequency and space.
To test this hypothesis, we look for preferred frequencies and structures in forcing. 
The nonlinear forcing $\hbff = \hbfu \nabla \hbfu$ is computed at each time step and SPOD is performed on this vector with the same parameters as . 
Figure~\ref{fig:forcing_SPOD_spectrum} shows the SPOD spectrum of the forcing for $m=0$ and the real part of leading forcing structure at $\mathrm{St}=0.48$.
The forcing spectrum in figure~\ref{fig:forcing_SPOD_spectrum}(a) shows two sharp peaks at $\mathrm{St}=0.48$ and $\mathrm{St}=0.96$. Apart from this, no clear separation or amplification is visible. To illustrate the correlation of the forcing, we plot the shape of the forcing before the sharp peak at $\mathrm{St}=0.45$ [see figure~\ref{fig:forcing_SPOD_spectrum}(b)], and at the peak $\mathrm{St}=0.48$ [see figure~\ref{fig:forcing_SPOD_spectrum}(c)].
Before the sharp peak, the forcing field contains no discrete coherent structure and exhibits low gain separation, while it is highly structured and low-rank at the peak. The forcing is thus mostly white noise, except at the vortex shedding harmonics.
This indicates that the discrete shear-layer eigenmode at $\mathrm{St}=0.5$ provides a natural receptivity that structured forcing can couple into, sharpening the output spectrum.
As white-noise forcing is assumed in the RA framework, the damped shear-layer eigenmode is overshadowed by the non-modal amplification, resulting in a broadband peak in resolvent gains.
In order for the RA model to recover this peak, the coloured forcing would need to be considered as in the parasitic mode approach used for the transitional flow in \citet{symon_tale_2019}.

The low-rank behavior of the dynamics, as well as the strong alignment between the SPOD and resolvent modes motivate the low-order reconstruction of the second-order statistics.

\section{Low-order reconstruction of turbulent statistics}\label{sec:ROM}
In this section, we exploit the low-rank behavior of the SPOD and RA modes to build a low order model based only on the leading SPOD and RA mode. By using different scaling methods, we investigate how well the turbulent kinetic energy and turbulence-related WSS can be approximated. This will quantify the role of coherent structures on these quantities, which is of high importance for medical applications \cite{lantz2012quantifying, dyverfeldt2013magnetic}.

\subsection{Turbulent kinetic energy}
Turbulent kinetic energy (TKE) serves as a critical hemodynamic biomarker. Elevated near-wall TKE directly correlates with large fluctuations in wall shear stress, which drive endothelial dysfunction and vascular disease progression. 
The TKE characterizes the energy contained within the flow's velocity fluctuations and is defined by half the trace of the Reynolds stress tensor (RST).
By expressing the fluctuations through a basis of optimal resolvent response modes, we can efficiently reconstruct the RST and the resulting TKE fields.

Our goal is to reconstruct the RST. Due to the statistical homogeneity of the flow in the azimuthal direction $\theta$, the three-dimensional RST $\bfR_{ij}$ is axisymmetric. As detailed in Appendix~\ref{app:TKE}, nonlinear interactions between helical modes and their complex conjugate produce a non-zero contribution to the total axisymmetric RST. We can thus decompose the total RST into independent contributions from each azimuthal wavenumber $m$:
\begin{equation}
    \bfR_{ij} = \sum_{m\geq 0} \bfR_{ij, \, m}.
\end{equation}
The TKE for a specific azimuthal mode $m$ is then defined as $k_m = \bfR_{ii, \, m}/2$.

We restrict the reconstruction of stresses to structures in the frequency range $\mathrm{St} \in [0.1,1]$, where both the SPOD and RA spectra exhibit gain separation and the response is well approximated by the leading mode:
\begin{equation}
    \hbfu(\boldsymbol{x}, m, \omega) \approx \sigma^{(0)} \hbfu^{(0)}(\boldsymbol{x}, m, \omega) \cdot \underbrace{\left< \hbff^{(0)}(m, \omega), \hbff\right>}_{c(m, \omega)},
\end{equation}
where $\hbff$ is the true forcing. This rank-1 assumption is used to reconstruct the second-order statistics.

Due to the statistical homogeneity of the flow in the azimuthal direction $\theta$, the 3D RST $\bfR_{ij}$ 
is axisymmetric. For simplicity, we therefore refer to its $\theta$-average as the total RST.
Nonlinear interactions between helical modes and their complex conjugates produce a non-zero contribution to the total axisymmetric RST. Appendix~\ref{app:TKE} formally derives this relationship. The RST is reconstructed from the optimal resolvent response modes as
\begin{equation}\label{eq:RST_m}
    \bfR_{ij} \approx \sum_{m\geq 0} \bfR_{ij, \, m}^{(0)}
    , \quad \bfR_{ij, \, m}^{(0)}=
    \sum_{\omega} \Real \left( \sigma^{(0)}(m, \omega)^2 c(m, \omega)^2 \hat{u}_{i}^{(0)}(x, r, m, \omega) \hat{u}_{j}^{(0)^*}(x, r, m, \omega)\right).
\end{equation}
Evaluating half the trace of the RST then yields the TKE for the azimuthal mode $m$: $k_m = \bfR_{ii, \, m}$.

The expansion coefficients $c(m, \omega)$ are a priori unknown and need to be modeled for quantitative evaluation of the modes. The optimal scaling in the global sense scales the RA modes with the leading SPOD mode: $c(m, \omega) = \sqrt{\lambda^{(0)}} / \left(\sigma^{(0)} \langle \hbfu_{RA}^{(0)},\hbfu_{SPOD}^{(0)}\rangle \right)$ \cite{towne_spectral_2018}. This yields the best possible rank-1 reconstruction of the data using the optimal RA mode. However, this method requires SPOD modes derived from time-resolved data, which removes the predictive ability of the framework.
In contrast, assuming white-noise forcing with $c(m,\omega)=1$ scales the modes merely by their resolvent gains. This simple scaling is of particular interest for applications where time-resolved data are unavailable, such as 4D-flow MRI measurements. Although this does not recover the absolute magnitude of the stresses, it may capture the qualitative spatial distribution of the RST without time-resolved data.

\begin{figure*}
    \centering
    \def\svgwidth{1\textwidth}
    \fontsize{12}{1}\selectfont
\begingroup%
  \makeatletter%
  \providecommand\color[2][]{%
    \errmessage{(Inkscape) Color is used for the text in Inkscape, but the package 'color.sty' is not loaded}%
    \renewcommand\color[2][]{}%
  }%
  \providecommand\transparent[1]{%
    \errmessage{(Inkscape) Transparency is used (non-zero) for the text in Inkscape, but the package 'transparent.sty' is not loaded}%
    \renewcommand\transparent[1]{}%
  }%
  \providecommand\rotatebox[2]{#2}%
  \newcommand*\fsize{\dimexpr\f@size pt\relax}%
  \newcommand*\lineheight[1]{\fontsize{\fsize}{#1\fsize}\selectfont}%
  \ifx\svgwidth\undefined%
    \setlength{\unitlength}{720.33785122bp}%
    \ifx\svgscale\undefined%
      \relax%
    \else%
      \setlength{\unitlength}{\unitlength * \real{\svgscale}}%
    \fi%
  \else%
    \setlength{\unitlength}{\svgwidth}%
  \fi%
  \global\let\svgwidth\undefined%
  \global\let\svgscale\undefined%
  \makeatother%
  \begin{picture}(1,0.2030391)%
    \lineheight{1}%
    \setlength\tabcolsep{0pt}%
    \put(0.28529113,0.17840989){\color[rgb]{0,0,0}\makebox(0,0)[lt]{\lineheight{0}\smash{\begin{tabular}[t]{l}$m=0$\end{tabular}}}}%
    \put(0.7403614,0.18091332){\color[rgb]{0,0,0}\makebox(0,0)[lt]{\lineheight{0}\smash{\begin{tabular}[t]{l}$m=1$\end{tabular}}}}%
    \put(0.00010284,0.06977344){\color[rgb]{0,0,0}\makebox(0,0)[lt]{\lineheight{0}\smash{\begin{tabular}[t]{l}$k_{m, \, RA^{scaled}}$\end{tabular}}}}%
    \put(0.00004815,0.11773764){\color[rgb]{0,0,0}\makebox(0,0)[lt]{\lineheight{0}\smash{\begin{tabular}[t]{l}$k_{m, \, SPOD}$\end{tabular}}}}%
    \put(0.00004815,0.02188998){\color[rgb]{0,0,0}\makebox(0,0)[lt]{\lineheight{0}\smash{\begin{tabular}[t]{l}$k_{m, \, RA^{WN}}$\end{tabular}}}}%
    \put(0,0){\includegraphics[width=\unitlength,page=1]{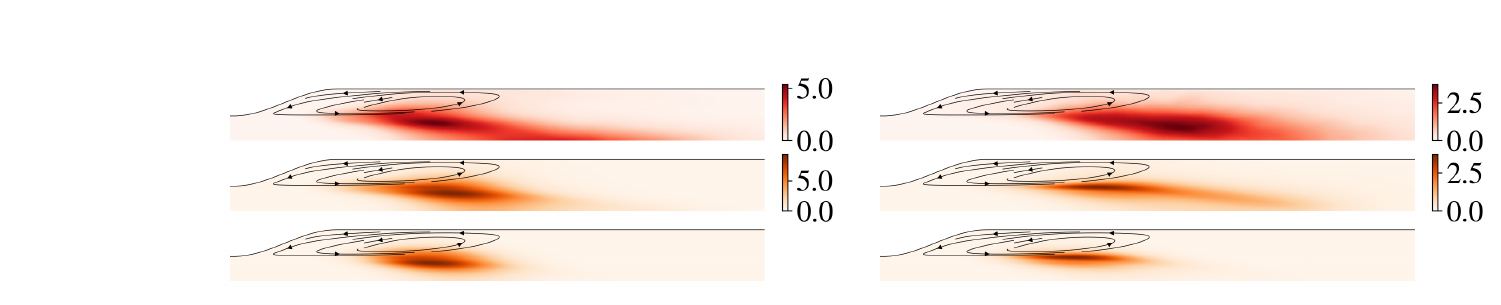}}%
  \end{picture}%
\endgroup%

    \caption{Turbulent kinetic energy reconstruction from fluctuations with wavenumbers $m=0,1,2$ from the SPOD leading mode, resolvent optimal response scaled with the SPOD gains and resolvent optimal response with white noise assumption. Values are normalised by $U_t^2 \times 10^{-3}$.}
    \label{fig:TKE_RA_SPOD}
\end{figure*}
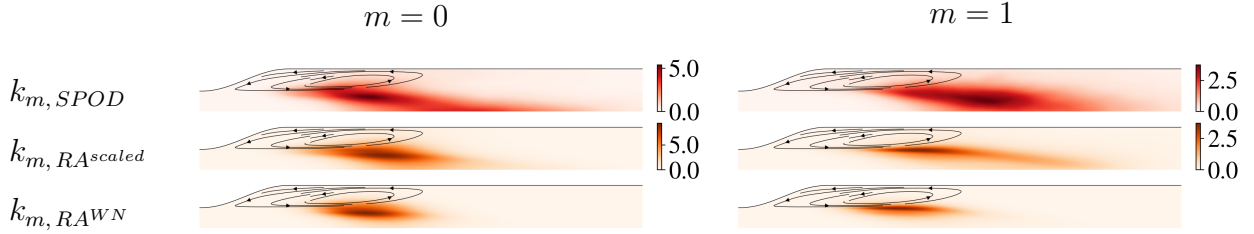

Figure~\ref{fig:TKE_RA_SPOD} compares the $k_m$ reconstruction for $m=0,1$ using the leading SPOD mode ($k_{m, \, SPOD}$), the optimally scaled leading resolvent response ($k_{m, \, RA^{scaled}}$), and the leading resolvent response assuming white noise forcing ($k_{m, \, RA^{scaled}}$).
For $m=0$, the optimally scaled RA-reconstruction of the TKE closely matches the SPOD result in the shear-layer. This was expected from the high gain separation and the dominance of the linear Kelvin–Helmholtz mechanism.  
For $m=1$, the optimally scaled RA only predicts TKE in a thin part of the shear-layer and misses most of the downstream turbulent dynamics.
This is expected from the weaker alignment for $m=1$ [see figure~\ref{fig:alignment}] and lower gain separation. 
When assuming white-noise forcing, the model still yields a satisfactory fit: it qualitatively captures the regions of high TKE, although it does not recover the absolute scaling. The model slightly underestimates the TKE downstream of the recirculation bubble. This discrepancy arises from differences in broadband amplification between the SPOD and the linear model. The resolvent gains place more weight on higher frequencies than the SPOD gains, which penalizes elongated modes and confines the TKE to the shear-layer near the recirculation bubble.
The successful TKE reconstruction in the shear-layer confirms that our linear model correctly captures the linear dynamics in this region. A better reconstruction in the breakdown region may be enabled by considering the true forcing and sub-leading RA modes.
Inferring the TKE from mean-flow analysis might offer an alternative method for medical application, particularly for the analysis of 4D-flow MRI data. 

\subsection{Turbulent wall shear stress} 

We investigate whether the accuracy of the second-order statistics reconstruction extends to relevant biomarkers through the turbulence-related WSS (tWSS), which is of primary interest in a hemodynamic context. For this it is very useful to first use the relation between the tWSS and the turbulent kinetic energy close to the wall which will be derived in the following.

A Reynolds decomposition is applied to the WSS vector, following \citet{lantz2012quantifying}:
\begin{eqnarray}
    \boldsymbol{\mathrm{WSS}} = \overline{\boldsymbol{\mathrm{WSS}}} + \boldsymbol{\mathrm{WSS}}' = \mu \left( \frac{\partial \babfu}{\partial \bfn} + \frac{\partial \bfu'}{\partial \bfn} \right)
\end{eqnarray}
where $\bfn$ denotes the wall-normal direction, and $\bfu$ the near-wall velocity normal to $\bfn$. 

For steady flows, the turbulent wall shear stress (tWSS) is defined as the norm of the root-mean-square (rms) of the fluctuating component $\boldsymbol{\mathrm{WSS}}'$ \cite{lantz2012quantifying}:
\begin{eqnarray}
    \mathrm{tWSS} &= \|\text{rms}(\boldsymbol{\mathrm{WSS}}')\|.
\end{eqnarray}
Assuming a linear velocity gradient in this region and neglecting the wall-normal velocity component close to the wall, the tWSS can be linked to the near-wall turbulent kinetic energy $k_{\delta}$ at a distance $\delta$ from the wall as
\begin{eqnarray}
    \mathrm{tWSS} = \frac{\mu}{\delta}\sqrt{2 k_{\delta}},
\end{eqnarray}
with $\mu$ the dynamic viscosity.
The derivation is detailed in Appendix~\ref{app:tWSS}.
This relation has been supported experimentally, showing a strong correlation between $\mathrm{tWSS}$ and $k_{\delta}$ observed in stationary flows \cite{ziegler2017assessment, andersson2017multidirectional}. 
The decomposition into azimuthal wavenumbers done for the RST in equations~\ref{eq:RST_m} can now be performed to the tWSS in a straightforward manner, reading:
\begin{eqnarray}\label{eq:sumtWSS}
    \mathrm{tWSS}^2 = \sum_{m\geq0} 2 \, \mathrm{tWSS}_m^2, \quad \mathrm{tWSS}_m^2 = 2\left ( \frac{\mu}{\delta} \right )^2 k_m,
\end{eqnarray}
where $\mathrm{tWSS}_m$ is the contribution from azimuthal wavenumber $m$ to the total $\mathrm{tWSS}$. Finally, the defined $\mathrm{tWSS} (x)$ is only a function of the axial position $x$.

\pgfplotstableread[col sep = comma]{tWSS_sqdimless.csv}\mytable
\definecolor{myblue}{RGB}{57, 106, 177}
\definecolor{myorange}{RGB}{218, 124, 48}
\definecolor{mygreen}{RGB}{62, 150, 81}
\definecolor{myred}{RGB}{204, 37, 41}
\definecolor{mypurple}{RGB}{107, 76, 154}

\begin{figure}
    \centering
    
    \begin{subfigure}[b]{1\textwidth}
        \centering
        \begin{tikzpicture}
            \begin{axis}[
                width=1.0\textwidth,
                height=0.3\textwidth,
                xmin=0, xmax=10,
                ymin=0, ymax=0.0019/1.9,
                ylabel={$\mathrm{tWSS}^2/(\mu D U^2)^2$},
                grid=minor,
                title style={at={(-0.10,1.05)}, anchor=north west, font=\bfseries},
                title={a)}
            ]
            
            \addplot[black, solid, ultra thick] table[x=x, y=tWSS_NL_m0] {\mytable};

            \addplot[myblue, solid, thick] table[col sep=comma, x=x, y=tWSS_t_m0]{\mytable};
            
            \addplot[myorange, solid, thick] table[x=x, y=tWSS_t_m1] {\mytable};
            
            \addplot[mygreen, solid, thick] table[x=x, y=tWSS_t_m2] {\mytable};
            
            \addplot[myred, solid, thick] table[x=x, y=tWSS_t_m3] {\mytable};
            
            \addplot[black, densely dotted, thick] table[x=x, y=upBlarge] {\mytable};\label{line:upB}
            
            \end{axis}
        \end{tikzpicture}
        \label{fig:tWSS_large_scale}
    \end{subfigure}

    \vspace{-0.9cm}
    
    \begin{subfigure}[b]{1\textwidth}
        \centering
        \begin{tikzpicture}
            \begin{axis}[
                width=1.0\textwidth,
                height=0.3\textwidth,
                xmin=0, xmax=10,
                ymin=0, ymax=0.0005/5,
                xlabel={$x/d_t$},
                ylabel={$\mathrm{tWSS}_m^2/(\mu D U^2)^2$},
                grid=minor,
                legend style={
                    at={(0.44, -0.35)}, 
                    anchor=north, 
                    legend columns=4, 
                    /tikz/column 1/.append style={column sep=10pt},
                    /tikz/column 2/.append style={column sep=10pt},
                    /tikz/column 3/.append style={column sep=10pt},
                    /tikz/column 4/.append style={column sep=10pt}
                },
                title style={at={(-0.10,1.05)}, anchor=north west, font=\bfseries},
                title={b)}
            ]
            
            
            \addplot[myred, solid, thick, forget plot] table[x=x, y=tWSS_t_m3] {\mytable};
            \addplot[myred, dashed, thick, forget plot] table[x=x, y=tWSS_SPOD_m3] {\mytable};
                    
            \addplot[mygreen, solid, thick, forget plot] table[x=x, y=tWSS_t_m2] {\mytable};
            \addplot[mygreen, dashed, thick, forget plot] table[x=x, y=tWSS_SPOD_m2] {\mytable};
        
            \addplot[myorange, solid, thick, forget plot] table[x=x, y=tWSS_t_m1] {\mytable};\label{line:tWSS_t_m1}
            \addplot[myorange, dashed, thick, forget plot] table[x=x, y=tWSS_SPOD_m1] {\mytable};\label{line:tWSS_SPOD_m1}
            \addplot[myorange, densely dotted, thick, forget plot] table[x=x, y=tWSS_RA_m1] {\mytable};\label{line:tWSS_RA_m1}
            
            \addplot[myblue, solid, thick, forget plot] table[col sep=comma, x=x, y=tWSS_t_m0]{\mytable};\label{line:tWSS_t_m0}
            
            \addplot[myblue, dashed, thick, forget plot] table[x=x, y=tWSS_SPOD_m0] {\mytable};\label{line:tWSS_SPOD_m0}
            \addplot[myblue, densely dotted, thick, forget plot] table[x=x, y=tWSS_RA_m0] {\mytable};\label{line:tWSS_RA_m0}

            \addlegendimage{black, solid, ultra thick}
            \addlegendentry{Total $\mathrm{tWSS}^2$}
            \addlegendimage{empty legend} \addlegendentry{}
            \addlegendimage{empty legend} \addlegendentry{}
            \addlegendimage{empty legend} \addlegendentry{}
            
            \addlegendimage{myblue, solid, thick}
            \addlegendentry{$\mathrm{tWSS}_{m=0}^2$}

            \addlegendimage{myorange, solid, thick}
            \addlegendentry{$\mathrm{tWSS}_{m=\pm1}^2$}

            \addlegendimage{mygreen, solid, thick}
            \addlegendentry{$\mathrm{tWSS}_{m=\pm2}^2$}

            \addlegendimage{myred, solid, thick}
            \addlegendentry{$\mathrm{tWSS}_{m=\pm3}^2$}

            \addlegendimage{myblue, dashed, thick}
            \addlegendentry{$\mathrm{tWSS}_{m=0,\,\mathrm{SPOD}}^2$}

            \addlegendimage{myorange, dashed, thick}
            \addlegendentry{$\mathrm{tWSS}_{m=\pm1,\,\mathrm{SPOD}}^2$}

            \addlegendimage{mygreen, dashed, thick}
            \addlegendentry{$\mathrm{tWSS}_{m=\pm2,\,\mathrm{SPOD}}^2$}

            \addlegendimage{myred, dashed, thick}
            \addlegendentry{$\mathrm{tWSS}_{m=\pm3,\,\mathrm{SPOD}}^2$}

            \addlegendimage{myblue, densely dotted, thick}
            \addlegendentry{$\mathrm{tWSS}_{m= 0, \, RA^{scaled}}^2$}
            \addlegendimage{myorange, densely dotted, thick}
            \addlegendentry{$\mathrm{tWSS}_{m=\pm1, \, RA^{scaled}}^2$}
            \addlegendimage{empty legend} \addlegendentry{}
            \addlegendimage{empty legend} \addlegendentry{}
        
            \end{axis}
        \end{tikzpicture}
    \end{subfigure}

    \caption{Axial distribution of the squared turbulent wall shear stress (tWSS). Panel (a) shows the total $\mathrm{tWSS}^2$ 
    resulting from the sum of all azimuthal contributions $\mathrm{tWSS}_m^2$. The upper wall is shown for reference (\ref{line:upB}).
    Panel (b) zooms on each azimuthal contribution $\mathrm{tWSS}_m^2$ (solid lines) extracted from LES data for the first four $m$ and presents the reconstructions derived from the leading SPOD modes (dashed lines) and the resolvent optimal response mode (dotted lines).}
    \label{fig:tWSS_data}
\end{figure}
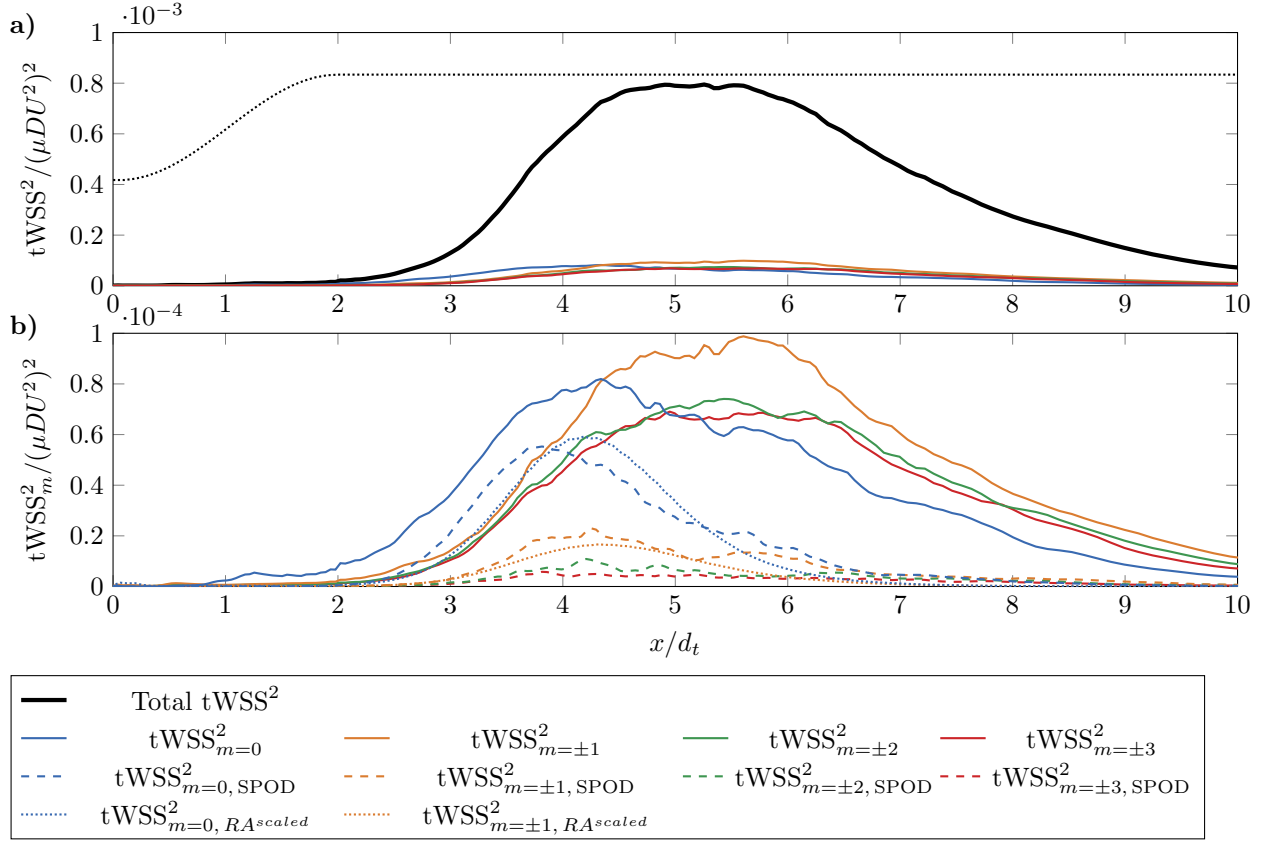

Figure~\ref{fig:tWSS_data}(a) shows the streamwise distribution of the total squared tWSS (black line) and $\mathrm{tWSS}_{m}$ the reconstruction of the tWSS using the $\theta$-decomposed fluctuations (plain color lines), computed exclusively from the full LES data.
The total tWSS peaks near $x/d_t = 5$, slightly upstream of the jet reattachment point at $x/d_t = 5.8$, which is in agreement with previous findings \cite{andersson2017multidirectional, andersson_characterization_2019}. 
Since all azimuthal contributions $\mathrm{tWSS}_m^2$ sum up to the total $\mathrm{tWSS}^2$ [see equation~\ref{eq:sumtWSS}], the squared values are plotted to allow direct visual assessment of the relative contribution of each azimuthal mode.
On average over $x/d_t \in [0, 5]$, the first three azimuthal modes account for more than $50\%$ of the total $\mathrm{tWSS}^2$, but as many as eleven modes are needed to reach 90\% of the total $\mathrm{tWSS}^2$.
Recovering the full magnitude of $\mathrm{tWSS}^2$ therefore requires multiple wavenumbers. Yet, some $m$ dominate in certain regions.

Figure~\ref{fig:tWSS_data}(b) compares the $\mathrm{tWSS}_m^2$ derived from the full LES dataset against the SPOD and resolvent low-order models.
It reveals that axisymmetric fluctuations (\ref{line:tWSS_t_m0}) generate the dominant tWSS contribution within the transitional subdomain ($x/d_t < 4.3$).
Further downstream ($x/d_t > 4.3$), sinuous fluctuations associated with $m=\pm1$ (\ref{line:tWSS_t_m1}) drive the highest stresses. This sinuous contribution peaks at the jet reattachment point ($x/d_t = 5.8$), where it accounts for $0.12\%$ of the total tWSS.

The dynamics at the wall can be considered low-rank in the regions where $\mathrm{tWSS}_{m, \, SPOD}$ accounting only for the leading SPOD mode (dashed lines) captures a significant portion of $\mathrm{tWSS}_m$ (solid lines).
For $m=0$, the leading SPOD mode (\ref{line:tWSS_SPOD_m0}) captures more than 50\% of $\mathrm{tWSS}_{m=0}$ (\ref{line:tWSS_t_m0}) in the region $x/d_t \in [2.9, 4.6]$.
For helical modes ($m\neq0$), the dynamics is of higher rank: for $m=1$ the leading mode alone carries a maximum of 36\% of the energy and 18\% for $m=2$.  

Figure~\ref{fig:tWSS_data}(b) also displays the reconstruction of the tWSS using the optimally scaled leading resolvent mode $\mathrm{tWSS}_{m, \, RA^{scaled}}$ (in dotted lines) for $m=0$ and $m=1$.
For both $m=0$ and $m=1$, the tWSS reconstructed from the optimal RA mode shows very good agreement with the leading SPOD mode in the shear-layer region ($x/d_t \in [0, 5]$). 
The $\mathrm{tWSS}_{m=0, \, RA^{scaled}}$ (\ref{line:tWSS_RA_m0}) peaks at $x/d_t = 4.1$, slightly downstream of the $\mathrm{tWSS}_{m=0, \, SPOD}$ peak ($x/d_t = 3.9$).
The $\mathrm{tWSS}_{m=1, \, SPOD}$ presents two peaks, one at $x/d_t = 4.2$ that is correctly captured by the model $\mathrm{tWSS}_{m=1, \, RA^{scaled}}$ (\ref{line:tWSS_RA_m1}), and a second one at $x/d_t = 5.8$ not captured. This shows that even under a white-noise forcing assumption, the RA reconstruction of the $\mathrm{tWSS}$ would capture the $\mathrm{tWSS}$ peaks in the shear-layer region.

By restricting the reconstruction to a selected frequency band, one can assess the impact of a given mechanism on the turbulent wall stresses. 
We evaluate the contribution of Kelvin-Helmholtz by limiting the $\mathrm{tWSS}$ reconstruction to the leading SPOD mode for $m=0$ over $\mathrm{St} \in [0.1, 1]$. This constrained rank-one reconstruction accounts, on average over the range $x/d_t \in [3, 5]$, for 81\% of the $\mathrm{tWSS}_{m=0}$ obtained from the leading SPOD mode summed over all available frequencies, and for 61\% of the $\mathrm{tWSS}_{m=0}$ reconstructed from all SPOD modes summed over all frequencies. 
This indicates that the Kelvin-Helmholtz instability has a substantial influence on the wall turbulent dynamics in the post-stenotic region. We note that the SPOD resolution in frequency does not allow to correctly evaluate the impact of the Coanda instability on the wall turbulent stresses. 

\section*{Conclusion}\label{sec:Conclusion}
This study investigates the transitional dynamics of a stenotic flow at $Re=4000$ and their impact on pathologically relevant hemodynamic biomarkers. To identify the underlying physical mechanisms, we use mean-flow resolvent analysis to model the dominant coherent structures and reconstruct turbulent statistics.
By comparing the physics-based resolvent predictions with data-driven SPOD modes extracted from LES, we assess the validity of a linear low-order framework for capturing the turbulent stresses. 

First, mean field analysis is used to explain the physical linear amplification mechanisms.
Global LSA identified a marginally unstable stationary mode ($m=1$) associated with a 
Coanda-type instability, as well as a discrete damped 
eigenmode at $m~=~0, \, \mathrm{St}~=~0.5$. This later eigenmode provides a natural receptive direction with which correlated forcing can couple, leading to a sharp peak in the SPOD spectrum.
Beyond this modal mechanism, resolvent analysis revealed a strong nonmodal amplification located in the shear-layer. Contrary to previous studies at lower Reynolds numbers, the most amplified fluctuations are found to be axisymmetric. Pseudospectral analysis reveals that this broadband amplification 
arises from non-normal interactions among several eigenvalues, confirming the convective nature of the linear mechanisms.
The high alignment between the optimal resolvent response and the leading SPOD modes in the shear-layer region confirmed that the dominant coherent structures are driven by linear mechanisms. 

Furthermore, we 
show that a rank-1 resolvent model can effectively reconstruct the leading second-order statistics in the bulk flow (TKE) and at the wall (turbulent WSS) using only the mean flow field. With optimal scaling based on SPOD information, the model accurately recovers the peak tWSS of the leading mode for $m=0$ and $m=1$. 
Even under a white-noise forcing assumption, 
the qualitative spatial distribution of these stresses was successfully captured.
However, reconstructing the total tWSS would require to sum contributions across the first eleven $m$ to reach 90\% of the total energy at the wall. 
%

This analysis is intended as a first step towards future application of mean-flow analysis of 4D-flow MRI data. The ability to estimate key flow dynamics and associated stresses solely from the mean flow opens possibilities for physics-informed diagnostic tools.

\begin{acknowledgments}
The research was supported by fundings from Elsa-Neumann
. The authors would like to acknowledge the valuable help of Dr. Hannes Dillinger for the LES setup.
\end{acknowledgments}

\appendix
\section{Linear operators linear analyses}\label{app:operators}
The resolvent operator is defined as
\begin{equation} 
\boldsymbol{\mathcal{R}} = \boldsymbol{\mathsf{P}}^\mathrm{T}(- i\omega \boldsymbol{\mathsf{B}} - \boldsymbol{\mathsf{L}})^{{-}1}\boldsymbol{\mathsf{P}}, 
\end{equation}
where the restrictor operators are
\begin{gather}
    \boldsymbol{\mathsf{B}} = 
        \begin{bmatrix} 1 & 0 & 0 & 0 \\ 0 & 1 & 0 & 0 \\ 0 & 0 & 1 & 0 \\ 0 & 0 & 0 & 0 \\ 
        \end{bmatrix}, 
        \quad
    \boldsymbol{\mathsf{P}} = 
        \begin{bmatrix} 1 & 0 & 0 \\ 0 & 1 & 0 \\ 0 & 0 & 1 \\ 0 & 0 & 0 \\ 
        \end{bmatrix}.
\end{gather}
The 2D linearized Navier–Stokes operator writes in cylindrical coordinates
\begin{gather}
\boldsymbol{\mathsf{L}} = \begin{bmatrix} \mathcal{A} +
\dfrac{\partial \overline{u}}{\partial x} & \dfrac{\partial
\overline{u}}{ \partial r} & 0 & \dfrac{1}{\rho}
\dfrac{\partial}{\partial x} \\ \dfrac{\partial
\overline{v}}{\partial x} & \mathcal{A} + \dfrac{\partial
\overline{v}}{\partial r} + \dfrac{\nu}{r^2} &
-\dfrac{2\overline{w}}{r} + \dfrac{2im\nu}{r^2} &
\dfrac{1}{\rho} \dfrac{\partial}{\partial r} \\
\dfrac{\partial \overline{w}}{\partial x} & \dfrac{\partial
\overline{w}}{\partial r} + \dfrac{\overline{w}}{r} -
\dfrac{2im\nu}{r^2} & \mathcal{A} +
\dfrac{\overline{v}}{r} + \dfrac{\nu}{r^2} & \dfrac{1}{\rho}
\dfrac{im}{r} \\ \dfrac{\partial}{\partial x} &
\dfrac{\partial}{\partial r} \dfrac{1}{r} &
\dfrac{im}{r} & 0 \end{bmatrix},
\end{gather}
where
\begin{equation} 
\mathcal{A} = \overline{u} \frac{\partial}{\partial x} + \overline{v} \frac{\partial}{\partial r} + \overline{w} \frac{im}{r} - \nu \left( \frac{\partial^2}{\partial x^2} + \frac{\partial^2}{\partial r^2} + \frac{1}{r}\frac{\partial}{\partial r} - \frac{m^2}{r^2} \right). 
\end{equation}

\section{Derivation of the Reynolds Stress Tensor from azimuthal contributions}\label{app:TKE}
The Reynolds stress tensor (RST), denoted $\bfR_{ij}$, is defined as the time-averaged second-order moment of the velocity fluctuations. Using the cylindrical coordinate system $\bm{x}=(x,r,\theta)$, the 3D RST reads:
\begin{equation}
    \bfR_{ij}(x,r,\theta)=\overline{u_i'(x,r,\theta,t)\,u_j'(x,r,\theta,t)},
\end{equation}
where the overline denotes time averaging.
We exploit statistical homogeneity in the azimuthal direction $\theta$ and compute the $\theta$-averaged RST (denoted by $\langle \bfR_{ij}\rangle_\theta$). We introduce the azimuthal decomposition of the velocity fluctuations:
\begin{equation}
    u_i'(x,r,\theta,t) = \sum_{m \geq 0} \hat{u}_{i}(x,r,t,m) e^{im\theta} + \text{c.c.}.
\end{equation}

Substituting this into the definition of $\langle \bfR_{ij} \rangle_\theta$ and expanding the products yields:
\begin{align}
    \langle \bfR_{ij}(x,r)\rangle_\theta
    &= \Bigg[
        \Bigg\langle \overline{\Bigg( \sum_{m_1 \geq 0} \hat{u}_{i}(x,r,t,m_1)\, e^{i m_1 \theta} + \text{c.c.} \Bigg)
        \Bigg( \sum_{m_2 \geq 0} \hat{u}_{j}(x,r,t,m_2)\, e^{i m_2 \theta} + \text{c.c.} \Bigg)} \Bigg\rangle_\theta
    \Bigg] \\
    &= \Bigg[
        \overline{\hat{u}_{i}(x,r,t,0)\,\hat{u}_{j}(x,r,t,0)} \nonumber\\
    &\qquad +
        \sum_{\substack{m_1, m_2 \geq 0 \\ m_1 + m_2 \neq 0}}
        \overline{\hat{u}_{i}(x,r,t,m_1)\,\hat{u}_{j}(x,r,t,m_2)}\,
        \Big\langle e^{i(m_1 + m_2)\theta} \Big\rangle_\theta \nonumber\\
    &\qquad +
        \sum_{\substack{m_1, m_2 \geq 0 \\ m_1 + m_2 \neq 0}}
        \overline{\hat{u}_{i}(x,r,t,m_1)\,\hat{u}_{j}^*(x,r,t,m_2)}\,
        \Big\langle e^{i(m_1 - m_2)\theta} \Big\rangle_\theta \nonumber\\
    &\qquad +
        \sum_{\substack{m_1, m_2 \geq 0 \\ m_1 + m_2 \neq 0}}
        \overline{\hat{u}_{i}^*(x,r,t,m_1)\,\hat{u}_{j}(x,r,t,m_2)}\,
        \Big\langle e^{i(-m_1 + m_2)\theta} \Big\rangle_\theta \nonumber\\
    &\qquad +
        \sum_{\substack{m_1, m_2 \geq 0 \\ m_1 + m_2 \neq 0}}
        \overline{\hat{u}_{i}^*(x,r,t,m_1)\,\hat{u}_{j}^*(x,r,t,m_2)}\,
        \Big\langle e^{-i(m_1 + m_2)\theta} \Big\rangle_\theta
    \Bigg].
\end{align}

The first and last sums vanish upon $\theta$-averaging. The second and third sums are complex conjugates. Using $\langle e^{i(m_1-m_2)\theta}\rangle_\theta = \delta_{m_1=m_2}$, these two sums become:
\begin{equation}
\sum_{m>0} \overline{\hat{u}_{i}(x,r,t,m) \hat{u}_{j}^*(x,r,t,m)} + \text{c.c.} = \sum_{m>0} 2 \overline{\hat{u}_{i}(x,r,t,m) \hat{u}_{j}^*(x,r,t,m)}
\end{equation}

Therefore, the $\theta$-averaged RST can be written as a sum of azimuthal contributions,
\begin{equation}
\langle \bfR_{ij}(x,r)\rangle_\theta
=
\overline{\hat{u}_{i}(x,r,t,0)\,\hat{u}_{j}(x,r,t,0)}
+
\sum_{m>0} 2\,\Real\!\left\{\overline{\hat{u}_{i}(x,r,t,m)\,\hat{u}_{j}^*(x,r,t,m)}\right\}.
\end{equation}

It is convenient to define modal RST contributions as:
\begin{equation}
\bfR_{ij, \, m}(x,r)=
\begin{cases}
\overline{\hat{u}_{i}(x,r,t,0)\,\hat{u}_{j}(x,r,t,0)}, & m=0,\\
2\,\Real\!\left\{\overline{\hat{u}_{i}(x,r,t,m)\,\hat{u}_{j}^*(x,r,t,m)}\right\}, & m>0,
\end{cases}
\end{equation}
so that:
\begin{equation}
\langle \bfR_{ij}(x,r)\rangle_\theta=\sum_{m\ge 0}\bfR_{ij, \, m}(x,r).
\end{equation}

From a physical perspective, each helical mode $m\neq0$ contributes to the axisymmetric second-order statistics through its interaction with its complex-conjugate counterpart $-m$, which yields the factor $2\,\Real\{\cdot\}$ in $\bfR_{ij, \, m>0}$.

Finally, the turbulent kinetic energy is obtained as the trace of the RST:
\begin{align}
k(x,r)&=\frac{1}{2}\,\langle \bfR_{ii}(x,r)\rangle_\theta,\\
&= \sum_{m\geq0} k_m, \quad k_m =
\frac{1}{2}
\begin{cases}
\overline{u_{ii}(x,r,0,t)^2}, & m=0,\\
2\,\overline{|u_{ii}(x,r,m,t)|^2}, & m>0,
\end{cases}
\end{align}

\section{tWSS correlation with the turbulent kinetic energy at the wall}\label{app:tWSS}
For stationary flows, the tWSS is defined as the norm of the root-mean-square (rms) of the fluctuating component $\boldsymbol{\mathrm{WSS}}'$, following \citet{lantz2012quantifying}:
\begin{equation}
    \mathrm{tWSS} = \|\mathrm{rms}(\boldsymbol{\mathrm{WSS}}')\| = \left\| \sqrt{\overline{ \mu^2 \left( \frac{\partial \bfu'}{\partial \bfn} \right)^2 }} \right\|,
\end{equation}
where $\bfn$ denotes the wall-normal direction, and $\bfu$ the near-wall velocity normal to $\bfn$.
Assuming a linear velocity gradient in this region, the fluctuating shear stress can be expressed as
\begin{eqnarray}
    \boldsymbol{\mathrm{WSS}}' = \mu \frac{(\bfu_{\delta}' - \bfu_0')}{\delta}
\end{eqnarray}
where $\bfu_0$ and $\bfu_{\delta}$ are the velocities at the wall and at a distance $\delta$ from the wall, respectively. 
Enforcing the no-slip condition ($\bfu_0 = 0$) yields
\begin{align}
    \mathrm{tWSS} = \frac{\mu} {\delta}\left\|\sqrt{\overline{\bfu_{\delta}^{'^2}}}\right\| 
\end{align}
We take $\delta$ small enough to assume the wall normal velocity component to be 0. The velocity can be expressed in a local basis $\bfu_{\delta} = (u_{1, \delta} \, u_{2, \delta} \, 0) $, such that:
\begin{align}
        \mathrm{tWSS} 
         &= \frac{\mu}{\delta} \sqrt{\overline{u_{1, \delta}^{'^2}} + \overline{u_{2, \delta}^{'^2}}}
\end{align}
In this local frame, the turbulent kinetic energy near the wall ($k_{\delta}$) is
\begin{eqnarray}
    k_{\delta} = \frac{1}{2}\left(\overline{u_{1, \delta}^{'^2}} + \overline{u_{2, \delta}^{'^2}}\right)
\end{eqnarray}
This assumption of 2D flow close to the wall provides a direct link between tWSS and the near-wall turbulent kinetic energy $k_{\delta}$ as
\begin{eqnarray}
    tWSS = \frac{\mu}{\delta}\sqrt{2 k_{\delta}}.
\end{eqnarray}
%

%

\end{document}